\documentstyle[12pt]{article}

\begin{document}

\begin{titlepage}
\title{Topology and Strings: Topics in $N=2$\thanks{Lectures
given by C. G\'omez at the Enrico Fermi Summer School, Varenna,
July 1994.}}

\author{C\'esar G\'omez\thanks{Instituto
de Matem\'aticas y F\'isica Fundamental, Serrano 123,
28006 Madrid (Spain).} \\{\it Departement de Physique
Theorique, Universit\'e de Gen\`eve}  \\
and \\ Esperanza
L\'opez$^{\mbox{\footnotesize{\dag}}}$
\\ {\it Th-Division, CERN, 1211 Gen\`eve 23, Switzerland} }

\date{}

\maketitle

\end{titlepage}

\vspace{5mm}

{\it Or l\`a, o\`u il n'y a point de parties, il n'y a ni
\'etendue,
ni figure, ni divisibilit\'e possible. Et ces Monades sont les
v\'eritables Atomes de la Nature et en un mot les Elements des
Choses.}
\vspace{2mm}

\hspace{5cm} {\it G.W. Leibnitz (La Monadologie-3)}

\vspace{1cm}

\subsection*{ Contents:}

1. $N\!=\!2$ Algebra and Topological Field Theory

\hspace{5mm} 1.1 $N=2$ Algebra and BRST-cohomology

\hspace{5mm} 1.2 Two Dimensional TFT's: Operator Formalism

\hspace{5mm} 1.3 Observables and Hodge-representatives

\hspace{5mm} 1.4 Twisting $N=2$ SCFT's

\hspace{5mm} 1.5 Deformations Preserving Topological
Invariance: Coupling Constants

\hspace{13mm} and $t{\bar t}$-equations

\hspace{5mm} 1.6 Landau-Ginzburg Description

\hspace{1cm} 1.6.1 Landau-Ginzburg Theories

\hspace{1cm} 1.6.2 Residue Formulae

\hspace{5mm} 1.7 Frobenius Manifolds

\hspace{5mm} 1.8 $t{\bar t}$-equations and Special Geometry

\vspace{3mm}

\noindent
2. Topological Strings

\hspace{5mm} 2.1 Topological Gravity and Gravitational
Descendents

\hspace{5mm} 2.2 Gravity and Landau-Ginzburg

\hspace{5mm} 2.3 Contact Terms and Gravitational Descendents

\hspace{5mm} 2.4 Integral Representation of Contact Terms

\hspace{5mm} 2.5 Gravity and the $t$-part of the
$t{\bar t}$-equations

\hspace{5mm} 2.6 Verlinde-Verlinde Contact Term Algebra: Pure
Topological Gravity

\hspace{5mm} 2.7 The gravitational Meaning of the
$t{\bar t}$-equations

\hspace{5mm} 2.8 $t{\bar t}$ Contact Term Algebra for
${\hat c}\!=\!3$

\hspace{5mm} 2.9 Holomorphic Anomaly: the Genus Zero Case

\hspace{5mm} 2.10 Higher Genus and Quantum Geometry

\hspace{5mm} 2.11 Final Comments

\newpage

\section{$N=2$ Algebra and Topological Field Theory}

\vspace{1cm}

\subsection{$N=2$ Algebra and BRST-cohomology}

\vspace{7mm}

Given the $N=2$ supersymetric algebra
\begin{eqnarray}
(Q^{\pm})^{2} & = & 0 \\
\{ Q^{+} , Q^{-} \} & = & H \nonumber
\end{eqnarray}
a topological field theory (TFT) can be defined by declaring one
of the two SUSY generators, let us say $Q^{+}$, to be a BRST-charge.
The physical Hilbert space $\cal H$ of the TFT is defined as the
BRST cohomology and the physical observables $\phi_{i}$ are
constrained by the symmetry requirement
\begin{equation}
[ Q^{+} , \phi_{i} ] = 0
\end{equation}
We can provide the Hilbert space $\cal H$ with an inner product
$\langle \hspace{3mm} , \hspace{3mm} \rangle$ such that the
adjoint of $Q^+$ is $Q^-$. This allow
us to associate with each cohomology class a
"Hodge-representative"\footnote{Hodge's theorem for compact
manifolds without boundary stablish that any p-form can be
uniquely decomposed as a sum of exact, co-exact and a harmonic
form. The harmonic form (3) is the Hodge-representative.}
satisfying
\begin{equation}
Q^{+}|i\rangle=Q^{-}|i\rangle=0
\end{equation}
{}From (1.b) we observe that this basis is
 one to one related to the vacuum states
\begin{equation}
H|i\rangle=0
\end{equation}
In these lectures we will mostly reduce our study to two dimensional
topological field theories \cite{W}, \cite{W2}. Physically,
topological invariance means that the only space-time dependence
of correlation functions will be on its topology, which in two
dimensions is simply given by the genus. Topological invariance
is certainly a much larger symmetry that the more familiar
conformal invariance, this however does not mean that all
topological field theories are massless or, equivalently,
with a traceless energy-momentum tensor. As we will see it is
possible to write down lagrangians which are manifestly
independent of the metric, and in this sense topological,
possessing dimensionful coupling constants. The renormalization
group can be directly applied to these topological theories. The
critical points of the renormalization group flow will define
topological conformal field theories, which will be
characterized by two chiral, $Q^{\pm}$ and ${\bar Q}^{\pm}$, N=2
algebras.

\vspace{1cm}

\subsection{Two Dimensional TFT's: Operator Formalism}

\vspace{7mm}

In two dimensions, a TFT can be nicely described using the
operator formalism \cite{OF}. Let $\cal H$ be the physical Hilbert
space
and let us choose as a basis the Hodge-representatives defined
by equation (3). Given now a generic Riemann surface $\Sigma_{g}$
of genus $g$ with $n$ punctures ${p_{1},...,p_{n}}$, the
operator formalism definition of the corresponding TFT will
consist in associating with these geometrical data a quantum
state $|\Sigma_{g};p_{1},...,p_{n}\rangle$ satisfying
\begin{eqnarray}
Q^{+} |\Sigma_{g};p_{1},...,p_{n}\rangle & = & 0  \\
\delta |\Sigma_{g};p_{1},...,p_{n}\rangle & = & Q^{+}
|\eta \rangle \nonumber
\end{eqnarray}
where by $\delta$ we mean any change of the metric and
the positions of the punctures. Condition (5.1) implies that
$|\Sigma_{g};p_{1},...,p_{n}\rangle \in {\otimes}^{n} \cal H$, and
condition (5.2) reflects the topological nature of the theory,
namely, any geometrical change is represented by $Q^{+}$-exact
forms and therefore all the geometrical dependence of the state
$|\Sigma_{g};p_{1},...,p_{n}\rangle$ can be mapped into the same
BRST-cohomology class. Hence we can associate with any genus $g$
and any number of punctures $n$ a Hodge-representative state
$|g,n\rangle$ as follows
\begin{equation}
|g,n\rangle = \sum_{i_{1},...,i{n}} C^{i_{1}...i{n}}_{g}
|i_{1}\rangle \otimes ... \otimes |i_{n}\rangle
\end{equation}
where we sum over the basis (3) of the physical Hilbert space
$\cal H$, and with the constants $C_{g}^{i_{1}...i{n}}$
depending only on the topological data, namely the genus and
the number of punctures. To define the theory reduces
now to fix these constants. In order to do it we will imposse,
as usual, consistency with sewing.

A topological sewing can be defined by two operations $*$ and
$\hat *$ such that
\begin{eqnarray}
|g,n\rangle & = & |g_1,n_1\rangle * |g_2,n_2\rangle ,
\hspace{1cm} n_1 + n_2 = n+2
\nonumber \\
|g,n\rangle & = & {\hat *} |g \! - \! 1,n \! + \! 2\rangle
\end{eqnarray}

Using (6), we can define the $*$-operation as follows
\begin{equation}
|g,n\rangle =
\sum_{i} ( \sum_{j} {{C_{g_1}}^{i_1 ... i_{n_{1} \! - \! 1}}}_{j}
{C_{g_2}}^{j \, i_{n_{1} \! + \! 2} ... i_{n_{1} \! + \! n_{2}}} )
|i_1\rangle \otimes ... \otimes |i_{n_1 + n_2}\rangle
\end{equation}
with
\begin{equation}
{{C_g}^{i_1 ... i_n}}_{j} \equiv
{C_g}^{i_1 ... i_n \, l} \eta_{lj}
\end{equation}
where we have introduced a "sewing metric" $\eta_{ij}$.
The $*$-operation can be analogously defined as follows
\begin{equation}
|g,n\rangle = \sum_{i}  \sum_{j} {{C_{g \! - \! 1}}^{i_1 ... i_n
\, j}}_{j} |i_1\rangle \otimes ... \otimes |i_n\rangle
\end{equation}

Using (8) and (10), we get the following type of sewing equations
\begin{equation}
{C_g}^{i_1 ... i_n} =  {{C_{g_1}}^{i_1 ... i_k}}_{j}
{C_{g_2}}^{j \, i_{k \! + \! 1} ... i_n} = \sum_{j}
{{C_{g \! - \! 1}}^{i_1 ... i_n j}}_{j}
\end{equation}

An inmediate consequence of sewing is that all constants
${C_g}^{i_1 ... i_n}$ can be written as products of the
elementary three point functions $C_{0}^{ijk}$. The sewing
equations (11) will be automatically fulfilled if the elementary
three point constants satisfy the associativity condition
\begin{equation}
\sum_{m} {{C_{0}}^{ij}}_{m} {C_{0}}^{mkl} =
\sum_{m} {{C_{0}}^{ik}}_{m} {C_{0}}^{mjl}
\end{equation}

The net result of the sewing construction is that a TFT
is completely determined by a set of constants ${C_{0}}^{ijk}$
and the sewing metric $\eta_{ij}$. In the previous
discussion, we have not considered the dependence of
${C_{0}}^{ijk}$ on the coupling constants of the theory.
Before entering into that problem, we will like to use the
previous formalism for the explicit construction of correlation
functions.

\vspace{1cm}

\subsection{Observables and Hodge-representatives}

\vspace{7mm}

Let us consider a physical observable $\phi_{i}$ satisfying
condition (2). As it is in general the case for local quantum
field theory, we would like to associate with this observable a
physical state, i.e. a BRST cohomology class and more in
particular a Hodge-representative in this class. This can be
done as follows. Let as take a hemisphere with the
field $\phi_{i}$ inserted on it at the point p. In this way we
obtain at the boundary a physical state $|i\rangle_{p}$ satisfying
$Q^{+} |i\rangle_p = 0$.
When we change the position of the insertion,
the state we will obtain will differ from the
former one by $Q^+$-exact forms. A simple way to project on the
Hodge-representative, will be by gluing the hemisphere to an
infinitely long cylinder with fixed perimeter $\beta$.
Using now relation (1.b), we can project on the harmonic
representative by taking the limit
\begin{equation}
|i\rangle = \lim_{T \rightarrow \infty} e^{-TH} |i\rangle_p
\end{equation}
The state $|i\rangle$ satisfy
\begin{equation}
Q^{+} |i\rangle = Q^{-} |i\rangle =0
\end{equation}

By the construction we have used \cite{CV},
the state $|i\rangle$ associated
with the observable $\phi_i$ will in principle depend on the
perimeter of the cylinder $\beta$. This statement can sound a
priori a bit strange. In fact if for different values of
$\beta$ we obtain different harmonic representatives we will be
in contradiction with the topological invariance as introduced
in equation (5), namely the difference of two harmonic forms is
not a $Q^+$-exact form and, on the other hand, a change of the
perimeter seems to be an innocent geometrical variation. What is
the solution to this puzzle? To get the solution we need to
understand the perimeter $\beta$ used to map physical observables
into Hodge-representatives as a renormalization group point
or scale. In this sense, changes of $\beta$ will produce in
general variations in the coupling constants\footnote{A more
detailed characterization of the RG in topological field
theories will be presented in section 1.6}.
Now the cohomology class is defined relative to $Q^{+}$ which will
depend explicitely on these couplings. Therefore changing
$\beta$ we will get, in general, different harmonic forms
in different cohomology classes. After this comment we can try
to connect the operator formalism construction presented in
section 1.2, with the definition of correlation functions for
physical observables.

By means of the sewing procedure we have reduced the problem of
defining the states $|g,n\rangle$ to that of defining a topological
metric $\eta_{ij}$ and the set of elementary three point
functions $C_{0}^{ijk}$. Our task will be now to get these
building blocks of the TFT directly from the algebra of
observables. Let us consider two physical observables $\phi_{i},
\phi_{j}$ inserted on the hemisphere, and let us project on a
Hodge-representative by gluing an infinite cylinder of fixed
perimeter $\beta$.
The state $|i,j\rangle_{\beta}$ obtained by this procedure
is by construction a physical state and can be projected on a
basis of $\cal H$
\begin{equation}
|i,j\rangle_{\beta} = \sum C_{ij}^{k} ({\beta}) |k\rangle_{\beta}
\end{equation}
The constants $C_{ij}^{k} (\beta)$ define the cohomology ring
structure for a particular set of Hodge-representatives, namely
the ones defined at the renormalization point
$\beta$.

Now we can use these cohomology ring constants, and
sewing, to define any correlator of physical observables
\begin{equation}
\langle \phi_i \phi_j \phi_k \phi_l
\rangle_{0} = \sum_{nm} C_{ij}^{n} \eta_{nm}
C_{kl}^{m} = \sum_n C_{ij}^{n} C_{nkl}
\end{equation}
{}From (16) we get in particular
\begin{equation}
\eta_{ij} = \langle \phi_i \phi_j \rangle_{0}
\end{equation}

It should be stressed that the sewing metric or topological
metric, $\eta_{ij}$ defined by two point correlators on the
sphere, does not coincide with the inner product of $\cal H$
relative to which the adjoint of $Q^+$ is $Q^-$.
The dependence on the renormalization group point $\beta$ of
these correlators should be constrained to satisfy, as usual in
quantum field theory, renormalization group equations
\begin{equation}
\frac{d}{d \beta} ( C_{g}^{i_1 ... i_n}) = 0
\end{equation}

It will be important for the rest of our study to have
control on the $\beta$-dependence of the Hodge-representative
states. To do that, it is convenient to pass from the
abstract discussion we are developping until now to some
concrete cases of TFT.

\vspace{1cm}

\subsection{Twisting N=2 Super Conformal Field Theories}

\vspace{7mm}

In the previous section we have presented the general structure of
a TFT. To materialize this structure in one concrete
case, we will define TFT's associated with $N=2$ super conformal
field theories (SCFT).

The chiral algebra of a $N=2$ SCFT \cite{Wa} is generated by the
identity,
the energy momentun tensor $T(z)$, two supersymmetric currents
$G^{\pm}$ and a ${\cal U}(1)$ current $J(z)$. In terms of the
corresponding Laurent expansions
\begin{eqnarray}
T(z)= \sum_n L_n z^{-n-2} , & J(z)= \sum J_n z^{-n-1} , &
G^{\pm} (z) = \sum_n G_{n}^{\pm} z^{-n-3/2}
\end{eqnarray}
The $N=2$ algebra is given by
\begin{eqnarray}
&& \{ G_{r}^{-}, G_{s}^{+} \} = 2 L_{r+s} - (r-s) J_{r+s} +
(c/3)(r^2 - 1/4) \delta_{r+s,0} \nonumber \\
&& [ L_m , L_n ] = (m-n) L_{m+n} + (c/12)m(m^2 - 1)
\delta_{m+n,0} \nonumber \\
&& [ L_n , G_{r}^{\pm} ] = (n/2 - r) G_{n+r}^{\pm} \; \; \;,
\hspace{1.5cm} [ L_n , J_m ] = -m J_{m+n} \\
&& [ J_m , J_n ] = (c/3)m \delta_{m+n,0} \; \; \; , \hspace{1.9cm}
[ J_n , G_{r}^{\pm} ] = \pm G_{n+r}^{\pm} \nonumber
\end{eqnarray}
where $r$ and $s$ are intergers or half intergers depending if
the representation is in the NS or R sectors.
The same holds for the antiholomorphic components ${\bar
G}_{n}^{\pm}$, ${\bar J}_n$ and ${\bar L}_n$.

In order to build a TFT we want, first,
to use one of the two SUSY currents to define a BRST charge. This
is not possible inmediately because the SUSY currents have spin
$3/2$ instead of $1$, as should be the case for defining a
BRST-charge.
Second, we need an energy-momentum tensor that can be
written as an exact form in order to implement topological
invariance.
The two things can be achieved by twisting \cite{W2,EY}
the theory, which consists
in changing the energy-momentum tensor $T(z)$ to $T^{t} (z)$,
defined by
\begin{equation}
T^{t} (z) = T (z) + \frac{1}{2} \partial J(z)
\end{equation}
This change in the energy-momentum tensor corresponds to couple
the ${\cal U} (1)$ current to a background gauge field equal to
half the spin connection. The net result of this background
field is to change the spin $s$ of any field of charge
$q$ to $s- q/2$. This is the effective change of the spin coming
from the holonomy contribution for a charge $q$ coupled to a
${\cal U} (1)$ gauge field equal to one half the spin connection.

After twisting, the SUSY current $G^{+}$ of positive charge
$q=1$ and spin $3/2$, becomes a one form and can be used to
define a BRST charge
\begin{equation}
Q^{+} = \oint G^{+} (z) dz
\end{equation}
Moreover, from the algebra relations (20) we get
\begin{equation}
T^{t} (z) = \oint G^{+} (w) G^{-} (z) dw
\end{equation}
which makes the twisted energy-momentum tensor (21) a
$Q^{+}$-exact form.

Therefore, by twisting the $N=2$ SCFT we have obtained a
topological conformal field theory with two BRST charges
\begin{equation}
Q^{+} = \int G^{+} (z) dz , \hspace{1cm}
{\bar Q}^{+} = \int {\bar G}^{+} (z) dz
\end{equation}
and a traceless energy-momentum tensor $T^{t} (z)$,
${\bar T}^{t} ({\bar z})$ which are, relative to $Q^{+}$ and
${\bar Q}^{+}$, exact forms.

Before leaving this section, let us note that we could have done
the
twist coupling the ${\cal U} (1)$ charge to minus one half of the
spin connection. In this case, the $G^-$ current becomes the BRST
charge. This can be done
independently in the left and right sectors.

\vspace{6mm}

{\bf Physical Hilbert Space and Observables: Chiral Ring
\cite{LVW}}

\vspace{2mm}
The Hilbert space of the original $N=2$ SCFT is, as usual in
CFT's, a direct sum of irreps of the chiral algebra. Each irrep
is associated with a primary field which represents the
observables of the theory and is characterized by a
weight $\Delta$ and a ${\cal U} (1)$ charge $q$.

As an example we can mention the case of $A_{k+1}$ minimal models.
The central extension in this case is given by
\begin{equation}
c= \frac{3k}{k+2}
\end{equation}
with $k$ an interger number, called level. The irreps fulfilling
the Hilbert space are parametrized by
\begin{eqnarray}
weights: & \Delta_{l,m} & = \frac{l(l+2)-m^{2}}{4(k+2)} \hspace{1cm}
l=0,...,k \; \; ; m= -l, -l\!+\!2,...,l\!-\!2,l \nonumber \\
{\cal U} (1) charges: & q_m & = \frac{m}{k+2}
\end{eqnarray}
Each of these irreps is associated with a primary field
$\phi_{l,m} (z)$. Denoting $|l,m\rangle$ the weight vector,
the map between observables and states is given by
\begin{equation}
|l,m\rangle = \phi_{l,m} (0) |0\rangle
\end{equation}

When we twist the theory, the Hilbert space of the $N\!=\!2$
SCFT collapses into $Q^{+}$-cohomology classes.
The best way to understand this truncation is by using a Coulomb
gas representation where the BRST charge of the $N=2$ SCFT
is defined in terms of the screening currents \cite{F}. By the
twist we modify the theory in such a way that the energy-momentum
tensor becomes, relative to this BRST charge, an exact form.

Our task now will be to associate a Hodge-representative to each
cohomology class and to define the corresponding physical
observables. As usual, we can take as Hodge-representative the
harmonic or vacuum forms of the twisted theory
\begin{equation}
Q^{+}|i\rangle=Q^{-}|i\rangle=0
\end{equation}
These Hodge-representatives are precisely the Ramond vacua of
the original $N=2$ SCFT\footnote{For the minimal models (25)-(27),
solutions to (28) correspond to $l=m$.}. In other words, each
cohomology class of the twisted theory has as
Hodge-representative a Ramond vacua of the untwisted theory. It
is well known that the NS and R realizations
of a $N=2$ superconformal algebra are connected by the
spectral flow transformation ${\cal U}_{1/2}$
\begin{eqnarray}
{\cal U}_{1/2} \; L_0 \; {\cal U}_{1/2}^{-1} & = & L_0 -
\frac{J_0}{2} + \frac{c}{24} \nonumber \\
{\cal U}_{1/2} \; J_0 \; {\cal U}_{1/2}^{-1} & = & J_0 -
\frac{c}{6} \\
{\cal U}_{1/2} \; G_{{\mp} 1/2}^{\pm} \; {\cal U}_{1/2}^{-1} & = &
G_{0}^{\pm} \nonumber
\end{eqnarray}
The Ramond vacuum states are defined by
\begin{eqnarray}
L_0 |i\rangle_R & = & \frac{c}{24} |i\rangle_R \nonumber \\
G_{0}^{+}|i\rangle_R & = & 0
\end{eqnarray}
It is easy to see, using (29), that NS w.v. satisfying
$\Delta=q/2$ are one to one related to Ramond vacua (30).
These NS w.v. are associated in the $N=2$ SCFT with local
primary field $\phi_i$ that verify
\begin{equation}
[ G_{-1/2}^{+} , \phi_i ] = 0
\end{equation}

Fields satisfying (31) are called chiral fields.
Summarizing, each
cohomology class of the twisted theory is associated with a
chiral primary field. Indeed, equation (31) corresponds to the
BRST-invariance condition in the twisted theory, and it can be
proved that any general chiral field can be decomposed
into the sum of a chiral primary field and a $Q^+$-exact one.
The consistency of the twisting procedure
requires that the operator product expansion for chiral
primary fields is, up to $Q^+$-exact
forms, another chiral primary field. This is in fact the case.
In the twisted theory and due to the fact that the
energy-momentum tensor is $Q^+$-exact we can, up to $Q^+$-exact
forms, reduce our study of the operator product $\phi_i (z)
\phi_j (w)$ between two chiral primary field to the "topological
limit" $z \rightarrow w$. In this limit and by ${\cal U} (1)$
charge
conservation, the only possibility is another chiral primary
field $\phi_k$ such that
\begin{equation}
\phi_i \phi_j = C_{ij}^{k} \phi_k \; \; , \hspace{1cm}
q_k = q_i + q_j
\end{equation}
We reobtain in this way the ring structure we have already
discussed in section 1.3. This ring of observables is known as the
chiral ring \cite{LVW}. Analogously, there exist an antichiral
ring of
observables when we choose $G^-$
to define a BRST charge, and an spectral flow transformation
${\cal U}_{-1/2}$ connecting antichiral fields with
Ramond vacua.

It is nice to see that the topological theory defined by the
twisting mechanism implements in a natural way the spectral flow
transformation. In fact if we define the Hodge-representative $|
i \rangle$ by inserting on the hemisphere the chiral field
$\phi_i$ and projecting on the zero energy sector by gluing an
infinitely long cylinder, the state we get at the boundary will
have charge $q_i \! - \! c/6$, where $c/6$ comes
from the contribution of the twist to the functional integral
representation of the state $|i \rangle$. Now from (29) we see
that this is precisely the charge of the state ${\cal U}_{1/2}
\phi_i |0 \rangle $ obtained from the NS sector by spectral flow.

The anomaly of the ${\cal U}(1)$ current generated by the twist
imposses the following
selection rule for non vanishing correlators
$\langle \phi_{i_1} ... \phi_{i_n} \rangle_g$
\begin{equation}
\sum_i q_i = {\hat c} (1-g) \; \; , \hspace{1cm} {\hat c} = c/3
\end{equation}
In particular the topological metric $\eta_{ij} =
\langle j | i \rangle$
defined for Hodge-representatives, i.e. Ramond vacua, will be non
vanishing only if
\begin{equation}
q_i + q_j = {\hat c}
\end{equation}
or in other words, when the sum of the Ramond charges is equal
to zero.

\vspace{1cm}

\subsection{Deformations Preserving Topological Invariance:
Coupling Constants and the $t{\bar t}$-equations}

\vspace{7mm}

Physical observables of an $N=2$ SCFT are associated with chiral
superfields, of components
\begin{equation}
\Phi_i = ( \phi_{i}^{(0)} (z,{\bar z}) , \phi_{i}^{(1)}
(z, {\bar z}),
{\bar \phi}_{i}^{(1)} (z,{\bar z}), \phi_{i}^{(2)}(z, {\bar z}) )
\end{equation}
where
\begin{equation}
\phi_{i}^{(2)} = \{ Q^- , [ {\bar Q}^- , \phi_{i}^{(0)} ] \}
\end{equation}
and therefore
\begin{equation}
[ Q^+ , \int_{\Sigma} \; \phi_{i}^{(2)} ] = 0
\end{equation}
Similarly, for antichiral fields ${\bar \phi}_{\bar i}$ we can
define, relative to the SUSY charge $Q^+$
\begin{equation}
{\bar \phi}_{\bar i}^{(2)} = \{ Q^+ , [ {\bar Q}^+ ,
{\bar \phi}_{\bar i}^{(0)} ] \}
\end{equation}
which trivialy implies
\begin{equation}
[ Q^+ , \int_{\Sigma} \; {\bar \phi}_{\bar i}^{(2)} ] = 0
\end{equation}
Moreover, for $Q^+$ being the BRST charge,
${\bar \phi}_{\bar i}^{(2)}$
becomes a pure BRST field.

Using (36) and (38) we define a deformed theory parametrized by
the coupling constants $(t_i , {\bar t}_{\bar i})$ as follows
\begin{equation}
{\cal L} (t_i , {\bar t}_i) = {\cal L}_{0}^{N=2}  +
\sum_i t_i \int_{\Sigma} \phi_{i}^{(2)}  + \sum_{\bar i}
{\bar t}_{\bar i} \int_{\Sigma} {\bar \phi}_{\bar i}^{(2)}
\end{equation}

This deformed theory can be transformed into a TFT again by the
twisting mechanism. Let us fix a set of values $(t_0 , {\bar
t}_0 )$ for the coupling constants. If some of the non vanishing
coupling constants correspond to relevant deformations, then the
theory defined by (40) will represent a massive deformation of
the $N\! = \! 2$ SCFT, ${\cal L}_{0}^{N=2}$. For these massive
deformations, the only conserved ${\cal U} (1)$ current
correspond to the fermion number current (the difference between
the left and right charges at the conformal point). The TFT at
this point in the space of couplings is obtained by twisting
with respect to the conserved fermion number current
\begin{equation}
{\cal L}^T  = {\cal L} (t_i , {\bar t}_i) \; + \; \frac{1}{2}
\int j \; A
\end{equation}
for $A$ the ${\cal U} (1)$ spin connection and $j$ the fermion
number current. The antitopological twist is defined by
\begin{equation}
{\cal L}^{T^{\ast}} = {\cal L} (t_i , {\bar t}_i) - \frac{1}{2}
\int j \; A
\end{equation}
Physical observables of (41) are associated with the chiral
fields $\phi_i$ and the ones of (42) with the antichiral fields
${\bar \phi}_{\bar i}$.

If the deformed theory (40) is massive, the $N=2$
algebra generated by $Q^{\pm}$, ${\bar Q}^{\pm}$ will contain
non vanishing central terms of the type
\begin{equation}
\{ Q^+ , {\bar Q}^+ \} = \Delta \; , \hspace{1cm}
\{ Q^- , {\bar Q}^- \} = {\bar \Delta}
\end{equation}
The $N=2$ algebra (1) is then defined by
\begin{equation}
Q_{\pm} = \frac{1}{\surd 2} ( Q^{\pm} + {\bar Q}^{\pm} )
\end{equation}
After the twisting (41) ((42)), $Q_+$ ($Q_-$) becomes the
corresponding BRST charges.

For each point $(t, {\bar t})$ in the coupling space, we have
defined a BRST charge $Q_+ $ and therefore we can
fiber the coupling space by the cohomology ring. The study of
this bundle will be the main task for the rest of this section.

Let $|i,t, {\bar t}; \beta \rangle$ to be the state defined by
inserting on the hemisphere the field $\phi_i$ and projecting on
a zero energy state by gluing the hemisphere to an infinitely
long cylinder of perimeter $\beta$. This correspond to use for
the hemisphere a metric $g \! = \! e^{\phi} dz d{\bar z}$ with
$\beta \! = \! e^{\phi}$.
Let us now introduce a set of "connections" $A_i$, $A_{\bar
i}$\footnote{Properly speaking these connections are defined by
\[ \langle {\bar k} | \partial_i - A_i | j \rangle = 0 \]
with $|{\bar k} \rangle$ the antiholomorphic basis. The
connection (45) is then defined by
\[A_{ij}^{k} = A_{ij {\bar k}} g^{{\bar k} k} \]
with $g^{{\bar k} k}$ the inverse of the hermitian metric
$g_{i{\bar j}} \! = \! \langle {\bar j} | i \rangle$.}
as follows
\begin{eqnarray}
\partial_{t_i} |j,t, {\bar t}; \beta \rangle & =  & A_{ij}^{k}
\; |k,t,{\bar t}; \beta \rangle \: + \: Q^+ \! - \! exact
\nonumber \\
\partial_{{\bar t}_i} |j,t, {\bar t}; \beta \rangle & =  &
A_{{\bar i}j}^{k} \;
|k,t,{\bar t}; \beta \rangle \: + \: Q^+ \! - \! exact
\end{eqnarray}
Therefore the covariant derivatives are given by
\begin{equation}
D_i = \partial_i - A_i \; \; , \hspace{2cm} {\bar D}_{\bar i} =
\partial_{\bar i} - A_{\bar i}
\end{equation}
Using the functional integral representation of
$|i,t,{\bar t} \rangle$ and interpreting
the partial derivative
$\partial_i$ as the insertion and integration over the
hemisphere of the operator $\phi_{i}^{(2)}$, we can conclude,
by contour deformation techniques and equation (37), that
$\partial_i |j,t, {\bar t}; \beta \rangle$ is also a
physical state. With the same techniques, it is easy to see that
\begin{equation}
A_{{\bar i} j}^{k} = 0
\end{equation}

Defining now
\begin{equation}
A_{ijk} = \langle k | \partial_i | j \rangle = A_{ij}^{l}
\; \eta_{lk}
\end{equation}
for $\eta_{lk}$ the topological metric, we can derive, by
standard
functional integral arguments, curvature equations for the
connections $A_i$.
{}From (47) we get
\begin{equation}
\partial_{\bar l} A_{ijk} = \partial_{\bar l} A_{ijk} -
\partial_{i} A_{{\bar l}jk}
\end{equation}
which admits the functional integral representation
\begin{equation}
\langle \phi_k \: (\int_{\Sigma_L} \! Q^+ {\bar Q}^+
{\bar \phi}_{\bar l}) \; | (\int_{\Sigma_R} \! Q^-
{\bar Q}^- \phi_i ) \; \phi_j \rangle -
\langle \phi_k \: (\int_{\Sigma_L} \! Q^- {\bar Q}^-
\phi_i) \; |( \int_{\Sigma_R} \! Q^+ {\bar Q}^+
{\bar \phi}_{\bar l}) \; \phi_j \rangle
\end{equation}
where $\Sigma_L$ and $\Sigma_R$ represent the two hemispheres
glued respectively to infinitely long cylinders of perimeter
$\beta$.
To compute the first component of (50), we contract the
SUSY currents. The result is
\begin{equation}
\langle \phi_k \: (\int_{\Sigma_L} \! {\bar \phi}_{\bar l}) \; |
( \int_{\Sigma_R}
\! \partial {\bar \partial} \phi_i) \; \phi_j \rangle
\end{equation}
which can be written as
\begin{equation}
- \langle \phi_k \: (\int_{\Sigma_L} \! {\bar \phi}_{\bar l} ) \;
| ( \oint_C \partial_n  \phi_i) \; \phi_j \rangle
\end{equation}
for $C$ the boundary of $\Sigma_R$ and $\partial_n$ the normal
derivative along the cylinder. At the boundary the state defined
by inserting $\phi_j$ is projected on a zero energy state $|j
\rangle$, therefore, and taking into account that $\partial_n
\phi_i \! = \! [H,\phi_i ]$, we can rewrite (52) as
\begin{equation}
- \langle \phi_k \int_{\Sigma_L} \! {\bar \phi}_{\bar l} \; | \;
H \oint_C \phi_i \; |j \rangle = - \int d \tau \;
\langle k | \oint_{C_{\tau}} \! \! \! {\bar \phi}_{\bar l} \,
\; H \oint_C \phi_i \; |j \rangle
\end{equation}
We have used that
the state obtained by inserting $\phi_k$ and
${\bar \phi}_{\bar l}$ on the left hemisphere, after gluing the
cylinder, is anhilated by H. The integral over $\tau$ in (53)
is over the length of the left cylinder, $T$. Moving
$H$ to the left in (53), we obtain
\begin{equation}
\int d \tau \langle k | \; ( \partial_{\tau} \oint_{C_{\tau}} \!
\! {\bar \phi}_{\bar l}) \; \oint_C \phi_i \; |j \rangle
\end{equation}
The integration in $\tau$ is now performed easily, getting
contributions from the boundaries at $\tau \!=\!0,T$. The
contribution at $\tau\!=\!T$ cancels with an identical one coming
from the second term in (50). Then, we are left with
\begin{equation}
- \int \langle k | \; \oint_{C_{\tau}} \! \! \!
{\bar \phi}_{\bar l} \; exp(\!-\!TH) \oint_C \phi_i \; |j \rangle
\end{equation}
where the propagation of ${\bar \phi_{\bar j}}$ along the
infinitely long left cylinder, explicited by the factor
$exp(\!-\!TH)$, has the usual effect of projecting into the ground
states. Therefore (55) can be written in terms of the structure
constants of the chiral ring
\begin{equation}
-({\bar C}_{\bar l} C_i )_{kj}
\end{equation}
Using the same arguments for the second term in (50),
we finally obtain
\begin{equation}
\partial_{\bar l} A_{ij}^{k} = \beta^2 [ \: C_i ,
{\bar C}_{\bar l} \: ]_{j}^{k}
\end{equation}
with $\beta$ the perimeter of the cylinder\footnote{The previous
derivation of the
$t{\bar t}$-equation (57) admits a more geometrical
interpretation in the following terms. For $\Sigma_R$ one can
consider fixed $\phi_i$ at the point 1 and reduce the integral
over $\phi_i$ to integrate the moduli $\tau^R \! \in \! [0,
\infty]$, $\phi^R \! \in \! [ 0,2\pi]$ (see section 2.4). The
same for the part $\Sigma_L$ where one will fix ${\bar
\phi}_{\bar j}$ at 1 and represent the integration over the
insertion of ${\bar \phi}_{\bar j}$ by the one of
$\tau^L \! \in \! [0, \infty]$, $\phi^L \! \in \! [ 0,2\pi]$.
These computations define two contact terms (see section 2.4).
The conmutator after interchanging $\phi_i$ and ${\bar
\phi}_{\bar j}$ gives equation (57). Notice the difference in
this construction with the definition of the 4-point amplitude
on the Riemann sphere where we only count with one moduli
parameter.}. Equation (57), first
derived by Cecotti and Vafa \cite{CV}, togheter with
\begin{equation}
[ D_i, D_j ] = [ {\bar D}_{\bar i}, {\bar D}_{\bar j} ]  =
[ D_i ,{\bar C}_{\bar j} ] = [ {\bar D}_{\bar i}, C_j ] =0
\end{equation}
\[ [ D_i, C_i ] = [ D_j, C_i ] \; \; , \; \; [ {\bar D}_{\bar i},
{\bar C}_{\bar j} ] = [ {\bar D}_{\bar j} , {\bar C}_{\bar i} ] \]
which can be deduced by similiar techniques as (57),
are known as the {\it $t{\bar
t}$-equations}. To contract the indices of the topological and
antitopological structure constants in (57) we use the metric
$g_{i {\bar j}}$ of the physical Hilbert space $\cal H$, namely
\begin{equation}
{\bar C}_{{\bar l} j}^{k} = g_{j {\bar n}} {\bar C}_{{\bar l}{\bar
m}}^{\bar n} g^{{\bar m} k}
\end{equation}
where
\begin{equation}
g_{i {\bar j}} = \langle {\bar j} | i \rangle
\end{equation}
with $\langle {\bar j} |$ the adjoint of the state $|j \rangle$.
Recall that for the inner product of {\cal H} introduced in
section 1.1, the adjoint of $Q^+$ is $Q^-$. The functional
integral representation of the state $|{\bar j} \rangle$ is
obtained, using the twisted lagrangian (42), by inserting on the
hemisphere the antitopological field ${\bar \phi}_{\bar j}$ and
projecting in the standard way on the zero energy
representative. Using this functional integral representation
we can interpret the metric tensor $g_{i {\bar j}}$ as a
topological-antitopological correlator on the sphere, where we
glue the two hemispheres through an infinitely long cylinder
with fixed perimeter $\beta$ and where we twist the theory with
$+ \frac{1}{2} w$ on the right hemisphere and
$- \frac{1}{2} w$ on the left, for
$w$ the spin connection. Notice that the correlator defined in
this way is not a topological correlator. Its geometrical
meaning can be derived as follows. From the definition of the
connection $A_i$ we derive
\begin{eqnarray}
D_i g_{j {\bar k}} & = & 0 \\
D_i \eta_{ij} & = & 0 \nonumber
\end{eqnarray}
{}From (61) and (47), we get
\begin{equation}
\partial_i g_{j {\bar k}} = A_{ij}^{l} g_{l {\bar k}}
\end{equation}
which means that $A$ is the connection of the metric $g$
\begin{equation}
A_{ij}^{l} = - g_{j {\bar k}} ( \partial_i g^{-1} )^{{\bar k} l}
\end{equation}
Using (63) we can rewrite the $t{\bar t}$-equation (57) as
equations for the metric $g$
\begin{equation}
\partial_{\bar l} ( g \partial_i g^{-1} )_{j}^{k} =
\beta^2 [ C_i , g {\bar C}_{\bar l} g^{-1} ]_{j}^{k}
\end{equation}

As we will see in section 1.8, the geometrical picture emerging
from these equations is closely connected with the special
geometry for special K\"ahler manifolds.

\vspace{1cm}

\subsection{Landau-Ginzburg Description}

\hspace{7mm}

Let us consider Landau-Ginzburg $N\!=\!2$ quantum field theories.
They are characterized by a superpotential $W$, the F-term,
which is a holomorphic function of $n$ chiral superfields $X_A$,
and a D-term defined by a K\"ahler potential $K(X_A, {\bar
X}_A)$. Using the canonical potential
\begin{equation}
K(X_A, {\bar X}_A) = \sum_{A=1}^{n} X_A {\bar X}_A
\end{equation}
the lagrangian reads
\begin{equation}
{\cal L} =  \int d^2 z d^4 \sum_{A=1}^{n} \theta X_A {\bar X}_A
+ \int d^2 z
d^2 \theta^+ W(X) + \int d^2 z d^2 \theta^- {\bar W}( {\bar X})
\end{equation}
Defining the superfields
\begin{eqnarray}
X_A & = & ( x_A, \psi_A , {\bar \psi}_A, F_A ) \\
{\bar X}_A & = & ( {\bar x}_A, \rho_A , {\bar \rho}_A, {\bar F}_A )
\nonumber
\end{eqnarray}
and after eliminating the F fields, using for that
the equations of motion
\begin{equation}
F_A = \frac{\partial {\bar W}}{\partial {\bar X}_A} \; \; ,
\hspace{1cm} {\bar F}_A = \frac{\partial W}{\partial X_A}
\end{equation}
we get in components\footnote{Here we assume that ${\bar W}$ is
the complex conjugate of $W$.}
\begin{equation}
{\cal L} = \int d^2 z ( - | \partial x_A |^2 +
\psi {\bar \partial} \rho_A + {\bar \psi}_A \partial {\bar
\rho}_A - | \partial_A W |^2
+ \partial_A \partial_B W \psi_A {\bar \psi}_B +
{\bar \partial}_A {\bar \partial}_B {\bar W} \rho_A {\bar \rho}_B)
\end{equation}

The F-term of lagrangian (66) is given by
\begin{equation}
{\cal L}^{F} =  \partial_A \partial_B W \psi_A {\bar \psi}_B
\end{equation}
and the \={F}-term by
\begin{equation}
{\cal L}^{\bar F} =  {\bar \partial}_A {\bar \partial}_B {\bar W}
\rho_A {\bar \rho}_B - | \partial_A W |^2
\end{equation}
with the rest defining the D-term.

After twisting the lagrangian ${\cal L}$, the fields $\psi_A$,
${\bar \psi}_A$ will become one forms and $\rho_A$, ${\bar
\rho}_A$ zero forms. This is, as usual, the net effect of the
coupling to a ${\cal U}(1)$ gauge field defined as $1/2$ of the
spin connection. Moreover, in the twisted theory the \={F} and
D-terms become BRST-exact forms and, therefore, we can define the
topological field theory by the F-term lagrangian (70). The
BRST-cohomology is given by \cite{VW,M}
\begin{equation}
{\cal R}_W = \frac{C [ X_A ]}{ [ W'(X_A) ] }
\end{equation}
i.e. the set of polynomials in the chiral superfields $X_A$
modulo the ideal generated by $W'(X_A)$.

What is known as the Landau-Ginzburg representation of a TFT is
to find a superpotential $W$ such that (72) coincides with the
chiral ring. Given a superpotential $W$ and a basis $\{
\phi_i(X_A) \}$ of ${\cal R}_W$, the ring structure constants
are defined by
\begin{equation}
\phi_i ( X  ) \; \phi_j ( X  ) =
C_{ij}^{k} \; \phi_k ( X ) \; mod W'
\end{equation}

It is important at this point to realize the different behaviour
under scale transformations of the world sheet metric
\begin{equation}
g \rightarrow \lambda^2 g
\end{equation}
of the F and \={F} lagrangians (70) and (71). While (70) is
invariant under transformation (74), the \={F}-term will
transform\footnote{The transformation law (75) in
the twisted theory comes from the fact that $\rho_A$, ${\bar
\rho}_A$ are, after twisting, zero forms.}
\begin{equation}
{\cal L}^{\bar F} \rightarrow \lambda^2 {\cal L}^{\bar F}
\end{equation}
Due to the invariance of (70), the scale transformations will
act as automorphisms
of the chiral ring (72). This is consistent
with the non-renormalization theorems for $N\!=\!2$ quantum
field theories. These theorems, which are mainly based on the
holomorphicity of the superpotential, imply that F-terms are not
corrected perturbatively and even non-perturbatively. Hence the
renormalization group transformation will preserve the chiral
ring structure (72) which only depends on F-terms.

Let us denote $|i,t,{\bar t}; \beta \rangle$ the state
associated with the observable $\phi_i (X_A)$. The functional
integral representation of this state can be formally written
like
\begin{equation}
|i, t ,{\bar t}; \beta \rangle = \int  \prod_{A=1}^{n} dX_A
d{\bar X}_A \phi_i(X) exp(\!-\! \int_H {\cal L}^{F})
exp(\!- \! \int_H {\cal L}^{\bar F})
\end{equation}
where the integration in the exponents is on the hemisphere $H$
used to define the state. The parameter $\beta$, as usual,
represents the perimeter of the hemisphere. The transformation
$g \rightarrow \lambda^2 g$ is now interpreted in two
complementary ways. First, it changes the non conformal part of
the lagrangian in the way described above. Secondly and based on
the non renormalization of the superpotential $W$, the change
$z \! \rightarrow \! \lambda z$, $\theta \! \rightarrow \!
\lambda^{-1/2} \theta$ amounts to a change $\int dz^2 d \theta^2
W \! \rightarrow \! \lambda \int dz^2 d \theta^2
W$ which can be compensated by changing
the couplings
$t_i$. This change of couplings in terms of the scale $\lambda$
would define the renormalization group $\beta$-functions for the
different couplings. Using these two facts and the equations
of motion for (69), we
obtain \cite{CV}
\begin{equation}
\beta^2 \frac{\partial}{\partial \beta^2}
|i,t,{\bar t};\beta
\rangle = -( \oint J_{0}^{5} + \frac{n}{2}) |i,t,{\bar t};\beta
\rangle + Q^+ \!-\!exacts
\end{equation}
with $\beta^2 \! \sim \! \lambda {\bar \lambda}$,
$\lambda$ in general complex.
The factor $\frac{n}{2}$ comes from the contribution of the
zero modes\footnote{The reader should be aware here that the
only non conformal piece of the lagrangian (69) is the
${\bar F}$-part.}.

At this point we can compose the previous computation with the
one we will perform for the twisted lagrangian (41). In that
case the $\beta$ dependence will come directly from the twist
term, which under dilatations transforms as
\begin{equation}
\int j \wedge d \phi \rightarrow \int j \wedge d (\phi + \epsilon)
\end{equation}
with $d \phi$ the spin connection for the metric $g_{z {\bar
z}}\!=\!e^{\phi} dz d{\bar z}$  \cite{CVV}.

{}From equation (77) we can easily obtain the dependence on
$\beta$ of the $t{\bar t}$-metric at the conformal
point\footnote{Recall that
for Landau-Ginzurg models ${\hat c}\!=\!\sum_{i\!=\!1}^{n}
(1\!-\!2q_i)$, with $q_i$ the charge of the chiral
field $X_A$ and $n$ the number of chiral fields. This
representation of $\hat c$ can be derived using singularuty
theory (see\cite{A}).}
\begin{equation}
g_{i {\bar i}} \sim (\beta^2)^{-q_i - \frac{n}{2}}
\end{equation}
with $q_i$ the Ramond charge of the state
$|i\rangle$.

We can now read equation (77) as defining a connection
$A_{\beta i}^{j}$. The $t{\bar t}$-equation for this connection is
\begin{equation}
\partial_{\bar l} A_{\beta i}^{j} = {\beta}^2 [ C_W ,
{\bar C}_{\bar l} ]_{i}^{j}
\end{equation}
where $C_W$ means multiplication by $W$ in ${\cal R}_W$.
At the conformal point we get $\partial_{\bar l}
A_{\beta i}^{j}\!=\!0$
since the quasi homogeneity of the superpotential implies
$C_W\!=\!0$.

\vspace{3mm}

\subsubsection{Landau-Ginzburg Representation}

\vspace{2mm}

In this subsection we will consider the problem of defining a
Landau-Ginzburg representation for the TFT defined by the lagrangian
\begin{equation}
{\cal L} = {\cal L}_{0}^{N=2} + \sum_{i} t_i \int \phi_{i}^{(2)}
\end{equation}
with ${\cal L}^{N=2}$ representing a twisted $N\!=\!2$ SCFT. We
will consider all ${\bar t}_i$-deformations equal to zero.

For a minimal $N\!=\!2$ SCFT at level $k$ (25)-(27), the chiral
ring is defined by
\begin{eqnarray}
\phi_i \phi_j & = & \phi_{i+j} \hspace{1cm} i+j \leq k \nonumber
\\
& = & 0  \hspace{12mm} i+j > k
\end{eqnarray}
This is isomorphic to the ring ${\cal R}_W$ for
\begin{equation}
W= \frac{X^{k+2}}{k+2}
\end{equation}
with only one chiral superfield $X$. The isomorphism is defined by
\begin{equation}
\phi_i \rightarrow X^i
\end{equation}

We consider now the deformed lagrangian (81)
and look for a superpotential $W(X,t)$ such that the
corresponding Landau-Ginzburg lagrangian is equivalent to it.
This in particular will mean that
\begin{equation}
\langle \phi_{i_1}(X,t) ... \phi_{i_s}(X,t) \rangle_{W(X,t)} =
\langle \phi_{i_1}(X) ... \phi_{i_s}(X) \rangle_{\cal L}
\end{equation}
for
\begin{equation}
\phi_i (X,t) = - \frac{\partial W}{\partial t_i}
\end{equation}
and where the l.h.s. of (85) is computed with the Landau-Ginzburg
F-term lagrangian\footnote{See subsection 1.6.2 on residue
formulae.} and the r.h.s. with the lagrangian (81).

Next we will follow the discussion in ref.\cite{DVV}
for determining $W(X,t)$. Let us assume
\begin{equation}
\phi_1 = X
\end{equation}
and define the perturbed ring structure as
\begin{equation}
X \phi_i = \phi_{i+1} + \sum_j a_{ij} t_{j-i+n+1} \phi_j
\end{equation}
where we assign ${\cal U} (1)$ charge $1-q_i$ to the couplings
$t_i$. The constants $a_{ij}$ are given by
\begin{equation}
a_{ij} = \langle \phi_1 \phi_i \phi_j \int \phi_{2k +1-i-j}^{(2)}
\rangle_0
\end{equation}

We can determine the value of $a_{ij}$ by the following
argument. For the perturbed theory defined by $t_j \!= \! 0 \;\;
\forall j \neq 1$, $t_1 \equiv t$, we get
\begin{eqnarray}
\phi_i \phi_j & = & \phi_{i+j} \hspace{12mm} i+j \leq k \nonumber \\
& = & t a_{ij} \phi_{i+j-k-1} \hspace{6mm} i+j > k
\end{eqnarray}
Impossing associativity to (90), with $a_{ij}$ again given by
(89), we obtain
\begin{eqnarray}
a_{ij} & = & 0 \hspace{12mm}  i+j \leq k \nonumber \\
& = & \mu \hspace{12mm} i+j > k
\end{eqnarray}
for $\mu$ some undetermined constant. Introducing this
solution into (88), we obtain polynomials in $X$ and $t$ for
representing the chiral fields $\phi_i$. The only thing that
remains now, is to get the superpotential with respect to which
(88) is the ring multiplication. A nice way to interpret (88) is
as diagonalizing the matrix $(C_1 )_{i}^{j}$ defined by the
multiplication rule, namely
\begin{equation}
\phi_1 \phi = {C_1}_{i}^{j} \phi_j = X \phi_i
\end{equation}
and therefore we can define $W$ by the characteristic equation
determining the eigenvalues of $(C_1)_{i}^{j}$
\begin{equation}
W'(X,t) = det (X \delta_{i}^{j} - {C_1}_{i}^{j}(t))
\end{equation}

This conclude the derivation of the superpotential associated
with the deformed lagrangian (81). The result however will
depends on the renormalization constant $\mu$. A change in the
scale $\mu$ can be represented by a change
$t_i \rightarrow \mu t_i$
in the couplings (see equation (88)). For a generic correlator
the dependence on $\mu$ will be $\mu^s$ with $s$ the number of
integrated fields entering into the correlator. If we also scale
the fields $\phi_i$ as $\phi_i \rightarrow \mu^{-\!q_i} \phi_i$,
we get an overall factor $\mu^{(-\!\sum q_i) + s}$.
By ${\cal U}(1)$-charge conservation, this factor is equal to
$\mu^{{\hat c}/2(2\!-\!2g)}$ and therefore can be cancelled
by introducing the string coupling constant coefficient
$\lambda^{2g-2}$ for $\lambda\!=\!\mu^{-{\hat c}/2}$.

\vspace{3mm}

\subsubsection{Residue Formulae}

\vspace{2mm}

Here we summarize the way to compute correlators in
Landau-Ginzburg theory. We will assume for the rest of this
section that $\bar W$ is the complex conjugate of $W$. We
consider the lagrangian
\begin{equation}
{\cal L} = {\cal L}^{D} + {\cal L}^{F} + {\tilde \lambda} {\cal
L}^{\bar F}
\end{equation}
where ${\cal L}^{D}$, ${\cal L}^{F}$ and ${\cal L}^{\bar F}$
were defined in section 1.6.
In the infrared limit ${\tilde \lambda}
\rightarrow \infty$ the main contribution to the Landau-Ginzburg
action comes from critical configurations \cite{V}
\begin{equation}
\frac{\partial W}{\partial X_A} = 0
\end{equation}
and the only contribution to the expectation values will come
from zero modes. For the bosonic part of ${\cal L}^{\bar F}$, we
get
\begin{equation}
\int \prod d X_A \; \;  exp( -{\tilde \lambda} | \partial_i W |^2 )
\end{equation}
which by gaussian integration around the critical points (95),
gives us
\begin{equation}
{\tilde \lambda}^{-n} ( H {\bar H})^{-1}
\end{equation}
with $H\!=\!det (\partial_A \partial_B W)$, the hessian of $W$.
The fermionic contribution contains two pieces, one from the
constant zero mode $\rho_{A}^{(0)}$ and the other from the $g$
holomorphic one forms $\psi_{A}^{(0)}$, if we are computing the
correlator in a genus $g$ Riemann surface. Hence the fermion
zero mode contribution is
\begin{equation}
{\tilde \lambda}^{n} H^g {\bar H}
\end{equation}
and therefore we get for the correlator
\begin{equation}
\langle \phi_{i_1} ... \phi_{i_s} \rangle_{W}^{g} =
\sum_{crit. points} \phi_{i_1}(X)...\phi_{i_s}(X) H^{g-1}
\end{equation}
where $\phi_{i_j}$ and $H$ are evaluated at the critical points.
Notice that the final result is ${\tilde \lambda}$-independent and
therefore we can use (99) as the general definition of
Landau-Ginzburg correlators. At genus $g\!=\!0$ and for a
"target space" of dimension one, we obtain
\begin{equation}
\langle \phi_{i_1}... \phi_{i_s} \rangle_{W}^{g=0} =
res (\frac{\phi_{i_1}...\phi_{i_s}}{\partial W}) \equiv
\frac{1}{2 \pi i} \oint_{\gamma} dX
\frac{\phi_{i_1}(X)...\phi_{i_s}(X)}{\partial W}
\end{equation}
with the contour $\Gamma$ going around the critical points of
$W$ \cite{V}, \cite{DVV}. Notice that at genus zero,
in order to get (100), we have
already integrated over the $\rho$-zero modes. The result (100) is
not invariant under the scaling $W \rightarrow \lambda W$.

\vspace{1cm}

\subsection{Frobenius Manifolds}

\vspace{7mm}

The concept of Frobenius manifolds \cite{D} is an useful
mathematical
tool for formalizing the structure of TFT's. Given a conmutative
and associative algebra A, with unity and non-degenerate
invariant inner product
\begin{equation}
\langle a, bc \rangle = \langle ab, c \rangle
\end{equation}
we will say that it is Frobenius if for a basis $e_i$
($i\!=\!1,...,n$) of A, the tensors $\eta_{ij}$ and $C_{ijk}$
defined by
\begin{eqnarray}
\langle e_i, e_j \rangle & = & \eta_{ij} \\
e_i e_j & = & C_{ij}^{k} e_k \nonumber
\end{eqnarray}
satisfy the following conditions
\begin{eqnarray}
\eta_{ij} & = & \eta_{ji} \nonumber \\
C_{ij}^{s} C_{sk}^{l} & = & C_{is}^{l} C_{jk}^{s} \\
C_{ijk} & = & C_{ij}^{l} \eta_{lk} \nonumber
\end{eqnarray}
and for $e\!=\!(e^i)$, the unit of A
\begin{equation}
e^s C_{sj}^{i} = \delta_{i}^{j}
\end{equation}
Notice that (103) are the generic conditions that we have
impossed
on the topological metric $\eta_{ij}$ and the ring structure
constants $C_{ijk}$ of a TFT.

A Frobenius manifold $M$ is a manifold which locally is a
Frobenius algebra. This means that at each point $x \in M$,
there exits tensors $\eta_{ij}(x)$, $C_{ij}^{k} (x)$ and unity
$e^i (x)$ satisfying conditions (103) and (104). We define the
invariant metric on $M$
\begin{equation}
d s^2 = \eta_{ij} dx^i dx^j
\end{equation}
relative to which the unit vector is covariantly constant.

For the metric (105), we can define local coordinates $t_i$
on $M$ such that \cite{D}
\begin{equation}
\eta_{ij} = cte
\end{equation}
We will call these coordinates coupling constants. The tensor
$C_{ijk}(t)$ in these coordinates satisfy the integrability
condition (see equation (58))
\begin{equation}
\partial_i C_{jkl} = \partial_j C_{ikl}
\end{equation}
which means that it can be represented
\begin{equation}
C_{ijk} = \partial_i \partial_j \partial_k F(t)
\end{equation}
with $F (t)$ being determined by the following set of equations
\begin{eqnarray}
\frac{\partial^3 F(t)}{\partial t_i \partial t_j
\partial t_k} \eta^{kl}
\frac{\partial^3 F(t)}{\partial t_l \partial t_m
\partial t_n} & = & \frac{\partial^3 F(t)}{\partial t_i
\partial t_m \partial t_k} \eta^{kl}
\frac{\partial^3 F(t)}{\partial t_l \partial t_j \partial t_m}
\nonumber \\
\frac{\partial^3 F(t)}{\partial t_i \partial t_j
\partial t_k} & = & C_{ijk}
\end{eqnarray}

For $\eta_{ij}$ the topological metric and $C_{ijk}$ the genus
zero three point function of a TFT it is easy to derive, by means
of standard Ward identities, equations (106)-(107) \cite{DVV},
using for such purppose the lagrangian
\begin{equation}
{\cal L} = {\cal L}^{(0)} + \sum_{i} t_i \int \phi_{i}^{(2)}
\end{equation}
Thus the coordinates $t_i$ can be identified with the coupling
constants in (110).

As an example we will consider the Frobenius manifold associated
with the Landau-Ginzburg superpotential for
minimal models (see section 1.6.1).
Let $M$ be defined by the following set of polynomials
\begin{equation}
M= \{ W(X, g_i) = X^{k+2} - (k+2) \sum_{i=0}^{k} g_i X^i \}
\end{equation}
The invariant inner product will be given by the residue formula
derived in the previous section
\begin{equation}
\langle f,g \rangle = res ( \frac{fg}{W'} )
\end{equation}

We can find the flat coordinates $t_i$ using the condition
$\eta_{ij}\!=\!cte$ and the inner product (112). The result is
\begin{equation}
t_i = - \frac{1}{k+1-i} res (W^{\frac{k+1-i}{k+2}})
\end{equation}
which defines the change from the Landau-Ginzburg coordinates
$g_i$ into the coupling constants $t_i$. We will come back to
equation (113) in a future section.

\vspace{6mm}

{\bf $t{\bar t}$-equations and Topological-Antitopological Fusion}

\vspace{2mm}
It is known \cite{D} that the $t{\bar t}$-equations can
be interpreted as the zero curvature condition for the system of
linear differential equations
\begin{equation}
\nabla_i \Psi = {\bar \nabla}_{\bar i} \Psi =0
\end{equation}
for
\begin{eqnarray}
\nabla_i & = & \partial_i + (g \partial_i g^{-1} ) - \lambda C_i \\
{\bar \nabla}_{\bar i} & = & \partial_{\bar i} - \lambda^{-1} {\bar
C}_{\bar i} \nonumber
\end{eqnarray}
with $\lambda$ a spectral parameter.

The mathematical meaning of
(114) and (115) as a way to fuse a topological and antitopological
theory was pointed out by Krichever in \cite{K}.
Given two topological theories characterized by $C_i$ and ${\bar
C}_{\bar i}$ respectively as ring structure constants, we  define
\begin{eqnarray}
( \partial_i - \lambda C_i ) \Phi = 0 \\
( \partial_{\bar i} - \lambda^{-1} {\bar C}_{\bar i} ) {\bar
\Phi} = 0 \nonumber
\end{eqnarray}
with $\Phi (t, \lambda)$ and ${\bar \Phi} (t, \lambda^{-1})$.
The essential singularities in (116) are in $\lambda\!=\!0$ and
$\lambda \! = \! \infty$. The $t{\bar t}$ fusion corresponds to
the Riemann-Hilbert problem of defining a functional $\Psi (
\lambda, t, {\bar t})$ such that at $\lambda \!=\!0$ behaves
like $\Phi$ and at $\lambda \!=\! \infty$ like $\bar \Phi$. The
solution to this problem is determined by equations (114) and
(115). The $t{\bar t}$-equations admit now the interpretation of
the isomonodromy equations for (114) and (115).

\vspace{1cm}

\subsection{$t{\bar t}$-equations and Special Geometry}

\vspace{7mm}

It is clear that the $t{\bar t}$-equations provide an extra
geometrical structure on the space of topological field
theories. In general the space of couplings constants
is a complex manifold with
coordinates $(t_i,{\bar t}_i)$\footnote{Notice the
difference between the Frobenius manifold defined
in the previous subsection, which contains only the couplings
$(t_i)$, and the full space of coupling constants
$(t_i,{\bar t}_i)$.}
and we can define in addition to
the topological metric $\eta_{ij}$ the hermitian metric
\begin{equation}
d s^2 = g_{i{\bar j}} dt^i d {\bar t}^{\bar j}
\end{equation}
In section (1.5) we introduced a connection $A_i$ such that both
$g_{i{\bar j}}$ and $\eta_{ij}$ are covariantly constant. This
connection can be written in terms of $g_{i{\bar j}}$ as in (62).
Moreover, $A_i$ defines a connection for the bundle obtained by
fibering the space of couplings with the BRST cohomology. The
$t{\bar t}$-equations satisfied by the connection $A_i$ are,
structuratly, very similar to the ones defining special K\"ahler
geometry \cite{SG}.
Before entering into a more detailed technical discussion , let
us try to undertand intuitively the physical origin of this
K\"ahler structure. The pieces we need for this discussion have
been already introduced in the previous sections and are
intimately connected with the meaning of renormalization group.

First of all, and reducing the discussion to Landau-Ginzburg
theories, we observe two interconnected phenomena

i) A reparametrization $W \rightarrow \lambda W$
in the superpotential induces a flow of the coupling constants.

ii) A world sheet reparametrization $g \rightarrow
\lambda g$ modifies the \={F}-part of the Landau-Ginzburg
lagrangian.
Recall that the \={F}-term is the non conformal part of the
twisted LG lagrangian.

\noindent
{}From i) and ii) follows that a rescaling $g \rightarrow \lambda
g$ of the world sheet metric induces both a change in $t$ and
$\bar t$ couplings. A point of view to understand the physical
meaning of the $t{\bar t}$-equation is as the lifting of this
renormalization group flow on the $t{\bar t}$ plane to the fiber
defined by the set of harmonic or zero energy states.
If now we think the "vacuum", the state of Ramond charge
$-{\hat c}/2$, as defining a line subbundle, i.e. we assume
conservation of charge, it is natural to translate the $(t{\bar
t})$ geometry into the characterization of the first Chern class
of the vacuum subbundle. Mathematically this picture will become
clear if the "$(t{\bar t})$-plane" defines a K\"ahler Hodge
manifold.

To check this intuitive picture requires to be able to define
the vacuum as a line bundle on the space of theories. This can
be done in a very particular case, namely when we are working on
the moduli space of a $N\!=\!2$ SCFT. In this case and due to
the independent left and right conservation of ${\cal U}(1)$
current we can decomposse the bundle defined by the BRST
cohomology into different charge sectors. Moreover we get the
constraint on the $t{\bar t}$-metric
\begin{equation}
g_{i{\bar j}} = 0 \hspace{1cm} if \; \; \; q_i + q_{\bar j} \neq 0
\end{equation}
Introducing the unit of the ring by
\begin{equation}
C_{i0}^{j} = C_{0i}^{j} = \delta_{i}^{j}
\end{equation}
we get from the $t{\bar t}$-equations, reduced to marginal fields
\begin{equation}
[ \partial_{\bar j} ( g \partial_i g^{-1} ) ]_{0}^{0} =
C_{i0}^{k} g_{k {\bar l}} {\bar C}_{{\bar j} {\bar 0}}^{\bar
l} g^{{\bar 0}0} = \frac{g_{i{\bar j}}}{g_{0 {\bar 0}}}
\end{equation}
Taking into account that
\begin{equation}
g_{0{\bar 0}} = \langle {\bar 0} | 0 \rangle
\end{equation}
we can use (120) to define a K\"ahler
potential and a Zamolodchikov metric \cite{Z} as
\begin{eqnarray}
G_{i{\bar j}} & \equiv & \frac{g_{i{\bar j}}}{g_{0{\bar 0}}} \\
K & = & - log \langle {\bar 0}|0 \rangle \nonumber
\end{eqnarray}
in such a way that
\begin{equation}
G_{i{\bar j}} = \partial_{\bar j} \partial_i K
\end{equation}
i.e. the standard definition of K\"ahler metric. Using (122)
we obtain the decompossition of the connection
$A_i\!= (\partial_i g) g^{-\!1}$ into two pieces.
The first is the K\"ahler connection on the moduli space,
defined as usual by
\begin{equation}
\Gamma_{ij}^{k} =  (\partial_i G_{j{\bar k}}) G^{{\bar k}k}
\end{equation}
and a second piece corresponding to the ${\cal U}(1)$ connection
of the line bundle generated by the vacuum
\begin{equation}
- \partial_i K
\end{equation}
which as usual for the Hodge-K\"ahler manifolds, is defined in
terms of the K\"ahler potential $K$. Comparing (79) (section
1.6) with (121) we observe that $K$ is
determined by the contribution of fermionic zero modes.

\vspace{3mm}

{\bf The ${\hat c}\!=\!3$ case and special geometry}

\vspace{2mm}

We will introduce first some generalities on special K\"ahler
manifolds. On a Hodge-K\"ahler manifold we introduce fields
$\phi_{p{\bar p}}$ of K\"ahler weight $(p,{\bar p})$ by the
following transformation rule
\begin{equation}
\phi_{p{\bar p}} \rightarrow \phi_{p{\bar p}} e^{-\frac{p}{2}f}
e^{-\frac{\bar p}{2} {\bar f}}
\end{equation}
with the ${\cal U}(1)$ gauge connection transforming like
\begin{equation}
\partial_i K \rightarrow \partial_i K + \partial_i f
\end{equation}
for $f$, $\bar f$ respectively holomorphic and antiholomorphic
functions. The covariant derivatives of these fields are defined
by
\begin{eqnarray}
D_i \phi_{p{\bar p}} & = & (\partial_i + \frac{p}{2} \partial_i K -
\Gamma_i ) \phi_{p{\bar p}} \\
{\bar D}_{\bar i} \phi_{p{\bar p}} & = & (\partial_{\bar i}
+ \frac{\bar p}{2} \partial_{\bar i} K ) \phi_{p{\bar p}} \nonumber
\end{eqnarray}

If $\phi_{p {\bar p}}$ is covariantly holomorphic
\begin{equation}
{\bar D}_{\bar i} \phi_{p {\bar p}} = 0
\end{equation}
Then we can define the holomorphic field ${\tilde \phi}_{p {\bar
p}}$ as
\begin{equation}
{\tilde \phi}_{p {\bar p}} = e^{\frac{\bar p}{2} K}
\phi_{p{\bar p}}
\end{equation}
which is a $(p-{\bar p},0)$ field. A Hodge-K\"ahler manifold is
special if there exits a symmetric tensor $W_{ijk}$ of K\"ahler
weight $(2,-2)$, such that \cite{SG}
\begin{eqnarray}
{\bar D}_{\bar l} W_{ijk} & = & 0 \\
D_i W_{jkl} & = & D_j W_{ikl} \nonumber \\
{R_{i_{\bar j}}}_{k}^{l} & = & G_{k{\bar j}} \delta_{i}^{l} +
G_{i{\bar j}} \delta_{k}^{l} - W_{ikn} {\bar W}_{{\bar j}
{\bar n}{\bar m}} G^{{\bar n}n} G^{{\bar m}l} \nonumber
\end{eqnarray}

Using (130), we can define a holomorphic tensor $C_{ijk}$ as
\begin{equation}
C_{ijk} = e^{-K} W_{ijk}
\end{equation}
which has weight $(4,0)$. From the integrability condition
(131.b), we can find a "covariant" prepotential $\hat S$
verifying
\begin{equation}
W_{ijk} = D_i D_j D_k {\hat S}
\end{equation}
with $\hat S$ again of K\"ahler weight $(2,-2)$. Analogously we
have
\begin{equation}
{\bar W}_{{\bar i}{\bar j}{\bar k}} = {\bar D}_{\bar i}
{\bar D}_{\bar j}{\bar D}_{\bar k} {\check S}
\end{equation}
with $\check S$ of weight $(-2,2)$. Defining the holomorphic
tensor ${\bar C}_{{\bar i}{\bar j}{\bar k}}$ of weight $(0,4)$ as
\begin{equation}
{\bar C}_{{\bar i}{\bar j}{\bar k}} \equiv e^{-K}
{\bar W}_{{\bar i} {\bar j}{\bar k}}
\end{equation}
and using $e^K {\bar D}_{\bar i} {\check S} \!=\! \partial_{\bar
i} e^K {\check S}$, we get
\begin{equation}
{\bar C}_{{\bar i}{\bar j}{\bar k}} = e^{2K} {\bar D}_{\bar i}
{\bar D}_{\bar j} \partial_{\bar k} S
\end{equation}
for
\begin{equation}
S \equiv e^K {\check S}
\end{equation}

The covariant prepotential $S$ allow us to integrate the special
geometry relation (131.c). Let us define
\begin{equation}
{\bar C}_{\bar i}^{l k} \equiv e^{2K} {\bar C}_{{\bar i}
{\bar l}{\bar k}} G^{{\bar l}l} G^{{\bar k} k}
\end{equation}
with the property
\begin{equation}
{\bar C}_{\bar i}^{lk} = \partial_{\bar i} S^{lk}
\end{equation}
where
\begin{equation}
S^{lk} = G^{{\bar l}l} \partial_{\bar l}
(G^{{\bar k} k}  \partial_{\bar k} S )
\end{equation}
Using this and the holomorphicity of $C_{ijk}$, we can write (131.c)
as
\begin{eqnarray}
\partial_{\bar j} \Gamma_{i k}^{l} & = & G_{k{\bar j}}
\delta_{i}^{l} + G_{i{\bar j}} \delta_{k}^{l}
- C_{ikn} {\bar C}_{\bar j}^{n l} = \\
& = & \partial_{\bar j} (  \partial_k K \delta_{i}^{l}
+ \partial_i K \delta_{k}^{l} - S^{l n} C_{ikn} ) \nonumber
\end{eqnarray}
Integrating (141) we obtain
\begin{equation}
\Gamma_{ik}^{l} =  \partial_i K \delta_{k}^{l}
+ \partial_k K \delta_{i}^{l} - S^{ln} C_{ikn} + f_{ik}^{l}
\end{equation}
with $f_{ik}^{l}$ a holomorphic tensor.

After this brief description of the special geometry, our next
task will be to identify the K\"ahler metric $G_{i{\bar j}}$
with the Zamolodchikov metric and the tensor $C_{ijk}$ with the
three point function. It is only in the particular case
${\hat c}\!=\!3$, where we have non vanishing
three point functions on the sphere for marginal fields, that
all the indices in (141) can be marginal. After
identiying the line bundle of the Hodge-K\"ahler manifold with
the one generated by the vacuum, we can define the state
$|0\rangle$ by a holomorphic section of weight $(2,0)$. In this
case the first Chern class can be defined in terms of the norm
of $|0\rangle$ as
\begin{equation}
\partial_i \partial_{\bar j} (log \| |0 \rangle \|^2) dz^i \wedge
d{\bar z}^j
\end{equation}
and we get
\begin{equation}
\langle {\bar 0} | 0 \rangle = e^{-K}
\end{equation}
in agreement with equation (122). Summarizing, in order to map the
$t{\bar t}$-geometry of the moduli space of ${\hat c}\!=\!3$
$N\!=\!2$ SCFT's into the special K\"ahler geometry we need to
identify the vacuum state with the trivializing holomorphic
section of the Hodge line bundle.

For Landau-Ginzburg theories we can write (144) as
\begin{equation}
\langle {\bar 0} | 0 \rangle = \int \prod_{A=1}^{n} dX_A
d {\bar X}_A \; exp(W-{\bar W})
\end{equation}
which makes explicit the $t$, $\bar t$ dependence.

\vspace{12mm}

\section{Topological Strings}

\vspace{5mm}

\subsection{Topological Gravity and Gravitational Descendents}

\vspace{7mm}

Topological gravity is the topological theory associated with the
moduli space of Riemann surfaces ${\cal M}_{g,n}$.
There are many good reviews on
this subject so we will concentrate the discussion on some
technical
points that will be relevant for our future analysis.

The aim of topological gravity is to get a topological field theory
representation of Mumford-Morita cohomology classes \cite{W3}.
Given a Riemann surface $\Sigma_{g,n}$ with genus $g$ and $n$
marked
points, we can consider its cotangent line bundle
at one of the points, namely $p_i$.
When we move the moduli
parameters of the surface, this defines a line bundle over
${\cal M}_{g,n}$. Let's denote by $\alpha_i$ the first Chern
class of this bundle.
The physical observables of topological gravity $\sigma_n (p_i)$,
called gravitational descendents,
are one to one related with the n-power of the two
form $\alpha_i$ in such a way that
\begin{equation}
\langle \sigma_{n_1} (p_1) ...\sigma_{n_s} (p_1) \rangle_g =
\int_{{\cal M}_{g,s}} \alpha_{1}^{n_1} \wedge ... \wedge
\alpha_{n}^{n_s}
\end{equation}
{}From (146) we see that
these amplitudes will be non vanishing only when it is fulfilled
the following selection rule
\begin{equation}
\sum_{i=1}^{s} n_i = 3g - 3 + n
\end{equation}

The simplest, from a physical point of view, way to realize this
topological field theory is to use the topological gauge theory
for the group ISO(2) \cite{VV}. After fixing the gauge,
the corresponding action is given by
\begin{equation}
S =  \int \pi \partial {\bar \partial} \phi + \chi \partial
{\bar \partial} \psi + b {\bar \partial} c + {\bar b} \partial
{\bar c} + \beta {\bar \partial} \gamma + {\bar \beta} \partial
{\bar \gamma}
\end{equation}
where $\phi$ is the Liouville field, $\psi$
its superpartner, $\pi$ and $\chi$ are Lagrange multipliers
conjugate to $\phi$ and $\psi$, and
$(b,c)$ and $(\beta , \gamma)$ are respectively the
ghost and superghost fields.
The ghost number assignations are the following:
zero for $\phi$, 1 for $\psi$, 1 for $c$ and 2 for $\gamma$.
The main ingredient to build a
topological field theory is the existence of a supersymmetric
charge $Q_S$ which behaves, BRST-improved, as an exterior
derivative
of the moduli space under study. It is
under $Q_S$ that all the fields are arranged into
supermultiplets.

For the action (148) the BRST charge is defined by
\begin{equation}
Q = Q_S + Q_g
\end{equation}
being $Q_S \! = \! \oint ( \partial \pi \psi + b \gamma)$
and $Q_g \!=\! \oint
c ( (T_L + \frac{1}{2} T_{gh} ) + \gamma ( G_L +
\frac{1}{2} G_{gh})$ respectively the $N\!=\!2$ and the
the gauge BRST charge. $T_L$ and $T_{gh}$ are the energy
momentum tensors of the Liouville and the ghost sectors, and
$G_L$, $G_{gh}$ the corresponding super stress tensors.
The topological nature of the action (148) is clear from the
following relations
\begin{eqnarray}
T_L & = & \{ Q_S , G_L \} \\
T_{gh} & = & \{ Q_S , G_{gh} \} \nonumber
\end{eqnarray}

Physical observables are defined by the BRST cohomology of (149).
In
topological gravity this cohomology turns out to be very simple.
In fact all physical observables are given by interger powers of
a field $\gamma_0$
\begin{equation}
\gamma_0 = \{ Q , \psi - {\bar \psi} \} = \frac{1}{2} \{ Q ,
\{ Q_S - {\bar Q}_S , \phi \} \}
\end{equation}
where $Q$ is the total BRST charge, i.e. left plus right.
Therefore we can define
\begin{equation}
\sigma_n = \gamma_{0}^{n}
\end{equation}
{}From (151) it looks that all observables of topological gravity
are BRST-trivial. The reason this is not the case is because we
are
interested in equivariant cohomology where we define the BRST
cohomology on gauge invariant objects. This doesn't include
$\gamma_0$, since $(\psi \! - \! {\bar \psi})$ is not a
gauge invariant quantity. We will come back to the discussion
on equivariant cohomology later.

The observables $\sigma_n$ given by (152) are zero forms on the
Riemann surface. We can define 1 and 2 forms by the following
recursive relations
\begin{equation}
d \sigma_n =  \delta_{BRST} \sigma_{n}^{(1)} \; \; , \hspace{1cm}
d \sigma_{n}^{(1)}  =  \delta_{BRST} \sigma_{n}^{(2)}
\end{equation}
The new operators $\sigma_{n}^{(1)}$ and $\sigma_{n}^{(2)}$ can
be integrated respectively over a one dimensional submanifold or
the whole surface $\Sigma$, giving also BRST invariant objects.

We want now to find a functional integral representation for
correlators $\langle \sigma_{n_1} ... \sigma_{n_s} \rangle_g $.
Instead of presenting the complete derivation we will,
qualitatively, motivate the final result.

The first thing will be to write the action (148) in a covariant
way.
The formulation of the $(\pi, \phi)$ system in (148) presents
problems
because the Liouville field $\phi$
behaves inhomogeneously under coordinate transformations.
In order to solve this, we can interpret $\phi$ as the conformal
factor for the metric $g\!=\! e^{\phi} \! {\hat g}$.
It can be shown that the physical quantities are independent of
the metric $\hat g$ chosen.
Under these conditions the conjugate
field $\pi$ gets coupled to the scalar curvature $R({\hat g})$.
Using $\int \sqrt{\hat g} {\hat R} \!=\! 2g \! - \! 2$, we must
cancell this background curvature by inserting operators
\begin{equation}
\prod_k e^{\alpha_k \pi(z_k)}
\end{equation}
in a set of points $\{z_k \}$ and in such a way that the
constants $\alpha_k$ satisfy
\begin{equation}
\sum_k \alpha_k = 2g-2
\end{equation}

Since the action (148) is supersymmetric, in order to
define a measure on ${\cal M}_{g,n}$, we need to integrate first
the
superpartners ${\hat m}_i$ of the moduli parameters $m_i$. The
integration of the supermoduli can be easily done because in
this rigid supersymmetry the supermoduli is split.
The integration of the $(3g \! - \! 3)$ odd moduli parameters
${\hat m}_i$ produces the insertion of super stress tensors
folded to Beltrami differentials $\chi_a$, ${\bar \chi}_{\bar a}$
\begin{equation}
\prod_{a, {\bar a} =1}^{3g-3} G (\chi_a) {\bar G}
({\bar \chi}_{\bar a})
\end{equation}
where $G \! = \! G_L \! + \! G_{gh}$.
In a similiar way, the integration over the superpartners of the
puncture moduli will produce insertions of the supertranslation
generator
\begin{equation}
\prod_{i,{\bar i}=1}^{s} \oint_{C_i} (b+G) \oint_{C_i} ({\bar b}
+{\bar G})
\end{equation}
with the contour $C_i$ defined around each puncture.

For the external states we must use
\begin{equation}
| \sigma_n \rangle = \sigma_n P |0 \rangle
\end{equation}
where $P\!=\!c{\bar c} \delta({\gamma}) \delta({\bar \gamma})$
is the puncture operator that reduces the number of
diffeomorphisms to those which leaves the puncture fixed.
Let's recall that the measure over ${\cal M}_{g,n}$ includes
the necessary factors to project out the zero modes of the ghosts
$b{\bar b}$ and $\delta({\beta})\delta({\bar \beta})$.
Due to this, when
we integrate the moduli of a puncture the operator $P$ is
reduced to $1$. Acting now with (157) on (158), we get as net
result for the external insertions $\sigma_n$ in the amplitudes
\begin{equation}
\prod_{i=1}^{s} \int_{\Sigma} \sigma_{n_i}^{(2)}
\end{equation}
where $\sigma_{n_i}^{(2)}$ is given by (153) and has ghost number
$n_i \! - \! 1$.

Combining now (154), (156) and (159) we obtain for the amplitudes
the following integral representation
\begin{equation}
\langle \sigma_{n_1} ... \sigma_{n_s} \rangle_g =
\int_{{\cal M}_{g,s}} \int e^{-S} \prod_{k} e^{\alpha_k \pi (z_k)}
\prod_{a,{\bar a} = 1}^{3g-3} G(\chi_a) {\bar G}({\bar
\chi}_{\bar a}) \prod_{i=1}^{s} \int_{\Sigma} \sigma_{n_i}^{(2)}
\end{equation}
By ghost number counting this will be non vanishing only if
\begin{equation}
\sum_{i=1}^{s} ( n_i -1 ) = 3g - 3
\end{equation}
in agreement with equation (147).

Let us next consider the coupling of topological matter
to topological gravity \cite{W3}. The gravitational
descendants $\sigma_n (\phi_i)$
associated with the chiral primary fields are simply defined by
\begin{equation}
\sigma_{n,i} = \phi_i \sigma_n \; \; , \hspace{1cm} n \geq 0
\end{equation}
Generic amplitudes $\langle \sigma_{n_1} (\phi_{i_1})...
\sigma_{n_s} (\phi_{i_s}) \rangle_g$ are again defined by equation
(160). The only additional information we need
to take into account is the extra ${\cal U} (1)$
background charge, modifying the selection rule (161) to
\begin{equation}
\sum_{i=1}^{s} (q_i + n_i -1)  =  3g-3 + {\hat c} (1-g)
\end{equation}
Therefore, we obtain
\begin{equation}
\langle \sigma_{n_1,i_1} ... \sigma_{n_s,i_s} \rangle_g =
\int_{{\cal M}_{g,s}} \int e^{-S} \prod_{k} e^{\alpha_k {\tilde
\pi} (z_k)}
\prod_{a,{\bar a} = 1}^{3g-3} G(\chi_a) {\bar G}({\bar
\chi}_{\bar a}) \prod_{j=1}^{s} \int_{\Sigma}
\sigma_{n_j,i_j}^{(2)}
\end{equation}
where ${\tilde \pi}$ is matter-modified conjugate of the
Liouville field.

The action $S$ in (164) is the unperturbed lagrangian ${\cal
L}_{0}^{N\!=\!2} \! + \! {\cal L}^{gr}$, with ${\cal L}^{gr}$
given by equation (148).
Our next task will be to generalize (164)
for a perturbed lagrangian. The approach will consist in
generalizing the Landau-Ginzburg description to topological
matter coupled to topological gravity.

\vspace{1cm}

\subsection{Gravity and Landau-Ginzburg}

\vspace{7mm}

In this section we will consider a TFT which posseses a
Landau-Ginzburg description in terms of a superpotential $W$,
coupled to topological gravity. This study will help in a better
understanding of some results derived in section 1.6.

To start with, we will first present a crucial theorem due to
Dijkgraaf-Witten. Let us consider the lagrangian general
perturbed lagrangian
\begin{equation}
{\cal L} = {\cal L}_{0}^{N=2} + {\cal L}^{gr} + \sum_{i=0}^{k}
t_i \int \phi_{i}^{(2)} + \sum_{n=1}^{\infty} \sum_{i=0}^{k}
t_{i,n} \sigma_n(\phi_i)^{(2)}
\end{equation}
where $t_i \! \equiv \! t_{i,0}$ and the small
phase space is defined by $t_{i,n}\!=\!0$, $n \geq 0$.
The identity operator, after coupling to gravity becomes the
puncture operator $P$.

Let us define the expectation values
\begin{equation}
u_i \equiv \langle P \phi_i \rangle \; , \hspace{2cm} i=0,...,k
\end{equation}
On the small phase space, we get
\begin{equation}
u_i = \eta_{ij} t_j
\end{equation}
Taking into account that $\eta_{ij}$ is invertible we can
interpret $u_i$ as a new set of coordinates on the small phase
space. Moreover, for any correlator computed on the small phase
space we obtain
\begin{equation}
\langle AB \rangle = R_{AB} (u_0,...,u_k)
\end{equation}
for some functional $R_{AB}$. The theorem proved in \cite{DW}
assures
that the correlator $\langle AB \rangle$ defined on the full
phase space, i.e. for the lagrangian (165) with $t_{i,n}\! \neq
\! 0$, is given by the same functional $R_{AB}$ where now the
coordinates $u_i$ (166) are computed taking into account the
couplings $t_{i,n}$. The proof of this theorem is based on the
topological recursion relations. What we need to show is that
\begin{equation}
\frac{\partial}{\partial_{i,n}} R_{AB} =
\frac{\partial u_l}{\partial t_{i,n}}
\frac{\partial R_{AB}}{\partial
u_l} = \langle \sigma_n (\phi_i ) AB \rangle
\end{equation}
Using the recursion relations for topological strings \cite{W3}
\begin{equation}
\langle \sigma_n ( \phi_i) AB \rangle = n \langle \sigma_{n-1}
(\phi_i) \phi_l\rangle \langle \phi^l AB \rangle
\end{equation}
we obtain
\begin{eqnarray}
n \langle \sigma_{n-1} (\phi_i) \phi_l \rangle \langle \phi^l AB
\rangle & = & n \langle \sigma_{n-1} (\phi_i) \phi_l \rangle
\frac{\partial R_{AB}}{\partial t_{l,0}} = \nonumber \\
= n \langle \sigma_{n-1} (\phi_i) \phi_l \rangle \langle \phi^l
P \phi_k \rangle \frac{\partial R_{AB}}{\partial u_k} & = &
\langle \sigma_n P \phi_k \rangle \frac{\partial
R_{AB}}{\partial u_k} = \\
& = & \frac{\partial u_k}{\partial t_{i,n}} \frac{\partial
R_{AB}}{\partial u_k} \nonumber
\end{eqnarray}
which concludes the proof of (168). The practical relevance of
this theorem is that allow us to get the form of the correlators
on the full phase space by doing a much simpler computation on
the small phase space.

As an illustrative example let us compute
the string anomalous dimension for the critical points of one
matrix models. We start with pure topological gravity, i.e. only
one primary field, the puncture $P$. The small phase space is
the complex line parametrizing the value of the cosmological
constant $t_0$. In the small phase space we get
\begin{equation}
u= \langle PP \rangle = t_0
\end{equation}
and for a generic correlator
\begin{equation}
\langle \sigma_i \sigma_j \rangle = \frac{1}{(i+j+1)!} \langle
\sigma_i \sigma_j P^{i+j+1} \rangle t_{0}^{i+j+1}
\end{equation}
Using the puncture equation \cite{DW}
\begin{equation}
\langle \sigma_i \sigma_j P \rangle = i \langle \sigma_{i-1}
\sigma_j
\rangle + j \langle \sigma_i \sigma_{j-1} \rangle
\end{equation}
we can rewrite (173) as
\begin{equation}
\langle \sigma_i \sigma_j \rangle = \frac{1}{i+j+1} u^{i+j+1}
\end{equation}
The previous theorem tell us that, on the full phase space, (175)
is the correct value for $\langle \sigma_i \sigma_j \rangle$ if
we replace $u$ by the value of $\langle PP \rangle$ on the full
phase space.

Taking into account all couplings we obtain
\begin{equation}
u= t_0 + \sum_{i=1}^{\infty} t_i \langle P P \sigma_i \rangle
\end{equation}
and from (175)
\begin{equation}
u = t_0 + \sum_{i=1}^{\infty} t_i u_i
\end{equation}
The $k$-th critical point \cite{GMK} is defined by $t_1 \!=\!1$,
$t_k\!=\!-1$, and from (177) we get
\begin{equation}
u = t_{0}^{1/k}
\end{equation}
The string anomalous dimension $\gamma$ is defined by
\begin{equation}
\langle 1 \rangle = t_{0}^{2-\gamma}
\end{equation}
Therefore, using (178) we have Kazakov's result
\begin{equation}
\gamma = - \frac{1}{k}
\end{equation}
Defining the string coupling constant $\lambda$ by
\begin{equation}
\langle 1 \rangle = \lambda^{-2}
\end{equation}
we obtain
\begin{equation}
\lambda^{-2} = t_{0}^{2+ 1/k}
\end{equation}

Dijkgraaf-Witten theorem underlines the equivalence of matrix
models and minimal topological strings. In fact, in the matrix
model
approach \cite{Do} we start with the KdV operator
\begin{equation}
L = D^{k+2} + (k+2) \sum_{i=0}^{k} v_i D^i
\end{equation}
with $D\!=\!\frac{d}{d X}$ for a formal parameter $X$. The KdV
flows are defined by
\begin{equation}
\frac{\partial L}{\partial {\tilde t}_p} =
[ (L^{\frac{p}{k+2}})_+,L]
\end{equation}
with $L_+$ the positive powers of $L$. Identifying
\begin{eqnarray}
v_k & = & \langle PP \rangle \\
{\tilde t}_p & = & t_{i,n} c_{i,n} \; \; , \hspace{1cm} p=
(k+2)n+i+1 \nonumber \\
c_{i,n}& = & ((i+1)(i+1+k+2)...(i+1+n(k+2)))^{-1} \nonumber
\end{eqnarray}
we obtain from (184)
\begin{equation}
\frac{\partial v_k}{\partial t_{i,n}} = c_{i,n} res (
L^{\frac{(k+2)n+i+1}{k+2}} )
\end{equation}
Denoting ${\hat L}\! \equiv \! L^{\frac{1}{k+2}}$ and
integrating $X$, we get
\begin{equation}
\langle P \sigma_n (\phi_i) \rangle = c_{i,n} res
( {\hat L}^{(k+2)n+i+1} )
\end{equation}
and similar relations for other correlators. From (187) we
observe how the correlator on the full phase space is defined by
a functional of the $(k+1)$ "coordinates" $v_k$ appearing in (183).

It is already clear the strong similarity of the matrix model
formula (187) and the residue formula we have derived in the
previous chapter for Landau-Ginzburg minimal models. Following
references \cite{DVV,L,EK}, we define the map from matrix models
into Landau-Ginzburg theories as follows
\begin{eqnarray}
{\hat L}^{k+2} & = & W \nonumber \\
\phi_i & = & [ {\hat L}^i \partial_X  {\hat L}]_+ \\
\sigma_n (\phi_i) & = & [ {\hat L}^{n(k+2)+i} \partial_X {\hat L}
]_+ c_{n-1,i} \nonumber \\
P & = & 1 \nonumber
\end{eqnarray}
which allows to represent the whole gravitational spectrum inside
the matter theory\footnote{For the extension of this map to
$W$-gravity see \cite{Le}.}.
Using this map we will now compare the matrix model expression
for correlators with the one we will obtain, from residue
formulae, in the Landau-Ginzburg framework.

For correlator $\langle \phi_i P \sigma_n (\phi_j) \rangle$, we
have in the matrix model formalism
\begin{equation}
\langle \phi_i P \sigma_n (\phi_j) \rangle =
\frac{\partial}{\partial t_i} \langle P \sigma_n (\phi_j)
\rangle = c_{j,n} \frac{\partial}{\partial t_i} res
( {\hat L}^{(k+2)n+j+1} )
\end{equation}
On the other hand, using (188) and the residue formulae, we get
in the Landau-Ginzburg formalism
\begin{eqnarray}
\langle \phi_i P \sigma_n (\phi_j) \rangle & = & c_{j,n-1} \int
(\frac{{\hat L}^{n(k+2)+j} \partial_X {\hat L}) \,
\phi_i}{W'} dX = \\
& = & c_{j,n-1} \int {\hat L}^{j+1 + (n-1)(k+2)} \,
\phi_i \, d X \nonumber
\end{eqnarray}
It is worth recalling that working in the Landau-Ginzburg
formalism, we are always at genus zero.
If we confine ourselves to the small phase space, we can use the
relation
\begin{equation}
\phi_i = \frac{\partial {\hat L}^{k+2}}{\partial t_i}
\end{equation}
{}From this we get
\begin{eqnarray}
\langle \phi_i P \sigma_n( \phi_j) \rangle & = & c_{j,n-1}
\int d X {\hat L}^{j+n(k+2)} \partial_i {\hat L} = \\
& = & c_{j,n} \frac{\partial}{\partial t_i}
res ( {\hat L}^{(k+2)n+j+1}) \nonumber
\end{eqnarray}
in agreement with (189). Notice that in principle (189) is a well
defined expression on the full phase space and the same for
(190) if we replace $\phi_i$ by $[{\hat L}^i \partial_X {\hat L}
]_+$, however, only on the small phase space we can use equation
(191). We will come back to this point in the next section.

\vspace{1cm}

\subsection{Contact Terms and Gravitational Descendents}

\vspace{7mm}

Let us consider the correlator $\langle \phi_i \phi_j \phi_k
\int \phi_{l}^{(2)} \rangle$ in Landau-Ginzburg theories. From
the residue formulae we obtain
\begin{eqnarray}
\langle \phi_i \phi_j \phi_k \int \phi_{l}^{(2)} \rangle & = &
\frac{\partial}{\partial t_l} \int \frac{\phi_i \phi_j \phi_k}{W'}
dX = \\
& = & - \int \frac{\phi_i \phi_j \phi_k {\phi'}_{l}^{2}}{{W'}^2}
d X +  \int \frac{1}{W'} [ (\frac{\partial \phi_i}{\partial
t_l}) \phi_j \phi_k + ...] \nonumber \\
\end{eqnarray}
Using (188) and (191) we get
\begin{equation}
\frac{\partial \phi_i}{\partial t_j} = \frac{\partial}{\partial
t_j} [ {\hat L}^i \partial_X {\hat L} ]_+ = [ \frac{\phi_i
\phi_j}{W'} ]_{+}^{'} \equiv C ( \phi_i, \phi_j)
\end{equation}
which are known as contact terms. From the matter representation
(188) of gravitational descendents we can compute the contact
term of a gravitational descendent with a primary field, so we
get in particular
\begin{equation}
C ( \sigma_n (\phi_i) , P ) = [ \frac{W' \! \int^X \! \!
\sigma_{n-1}
(\phi_i)}{W'} ]_{+}^{'} = \sigma_{n-1} (\phi_i)
\end{equation}
which is Saito's recursion relation \cite{S}. To derive (196) we
have used the following decomposition of $\sigma_n (\phi_i)$
\begin{equation}
\sigma_n (\phi_i) = W' \! \int^X \! \! \sigma_{n-1} (\phi_i) +
\sum_{l=0}^{k} a_l \phi_l = c_{i,n-1} [ {\hat L}^{(k+2)n+i}
\partial_X {\hat L} ]_+
\end{equation}
The part of $\sigma_n (\phi_i)$ projecting on chiral primary
fields can be easily obtained by noticing that $W' \! \int^X
\! \! \sigma_{n-1} (\phi_i)$ is a pure BRST object, namely
\begin{equation}
res_W (F+GW') = res_W (F)
\end{equation}
for any functions $F$ and $G$. In fact, from (198) we get
\begin{equation}
\langle \sigma_n (\phi_i ) \phi_j P \rangle = \sum_{i=0}^{k}
a_l \langle \phi_l \phi_j P \rangle
\end{equation}
and from (189)
\begin{equation}
\sum_{l=0}^{k} a_l \phi_l = \sum_{l=0}^{k} c_{n,i}
\frac{\partial}{\partial t_l}  res ( {\hat
L}^{(k+2)n+i+1}) \; \phi_{k-l}
\end{equation}
thus
\begin{equation}
\sum_{l=0}^{k} a_l \phi_l = \sum_{l=0}^{k}
\frac{\partial}{\partial t_l}
\langle P \sigma_n (\phi_i) \rangle \; \phi_{k-l}
\end{equation}

The contribution to correlators from the piece of
$\sigma_n(\phi_i)$ projecting on chiral fields originates
recursion relations. Let's take the three point function
\begin{eqnarray}
\langle \sigma_n (\phi_i) \phi_j \phi_k \rangle & = &
\sum_{l=0}^{k} \frac{\partial}{\partial t_l} \langle P
\sigma_n (\phi_i) \rangle \langle \phi_{k-l} \phi_j \phi_k
\rangle = \nonumber \\
& = & \sum_{l=0}^{k} \langle P \sigma_n (\phi_i) \phi_l \rangle
\langle \phi^{l} \phi_j \phi_k \rangle
\end{eqnarray}
while using (196) we obtain the recursion relation
\begin{equation}
\langle \sigma_n (\phi_i) \phi_j \phi_k \rangle
= \sum_{l=0}^{k} \langle \sigma_{n-1} (\phi_i) \phi_l \rangle
\langle \phi^{l} \phi_j \phi_k \rangle
\end{equation}

{}From the definition (195) it is clear that contact terms are
symmetric
\begin{equation}
C(\phi_i, \phi_j) = C ( \phi_j, \phi_i)
\end{equation}
Moreover, we can write
\begin{equation}
C( P_i, \phi_j ) \equiv A_{ij}^{k} P_k
\end{equation}
for $P_i$ either chiral primary or gravitational descendent.
Using again (195) we get
\begin{equation}
C(\phi_k , C(P_i, \phi_j)) = \frac{\partial A_{ij}^{l}}{\partial
t_k} P_l + A_{ij}^{l} C( \phi_k, P_l)
\end{equation}
as the rule for compossing contact terms.

The reader should notice that the Landau-Ginzburg description of
contact terms can not be trivially extended to the computation
of contact terms for two gravitational descendents $C(\sigma_n
(\phi_i), \sigma_m (\phi_j))$. The reason is again that we are
assuming relation (191) only on the small phase space.

\vspace{1cm}

\subsection{Integral Representation of the Contact Terms}

\vspace{7mm}

{}From the Landau-Ginzburg definition (195) of contact terms, it is
clear that the contribution to $C(\phi_i, \phi_j)$ is the
part in the product $\phi_i \phi_j$ which goes as $W'F$ for some
functional $F$. This is a priori a bit paradoxical taking into
account that for the matter theory, $W'F$ is a pure BRST-object
which decouples from any correlator. As it was first pointed out
by \cite{L,EK}, the reason for the
contribution of cohomologically trivial objects of the pure
matter theory to the contact terms is that they are non-trivial
in the equivariant cohomology which rules the spectrum after
coupling to gravity. To see this more clearly, let us introduce
the following integral representation of the contact terms
\begin{equation}
| C(\phi_i, \phi_j) \rangle = \int_D \phi_{i}^{(2)} |\phi_j\rangle
\end{equation}
where $D$ is an infinitesimal neigborhood of the point where the
operator $\phi_j$ is inserted.

A "sewing"\footnote{We will refer to "sewing"-representation
when the integration of a field over the Riemann surface is
transformed into integration over sewing parameters with all
punctures fixed.} or cancel propagator argument representation
of the contact term can be defined working with
$\phi_i$ and $\phi_j$ inserted on two fixed points of the
hemisphere
and integrating over the moduli $(\phi, \tau)$
with $ \phi \in [ 0,2\pi]$ and $\tau \in [ 0, \infty]$ as follows
\begin{equation}
|C(\phi_i, \phi_j) \rangle = \int_{0}^{\infty} d \tau
\int_{0}^{2\pi}
 d \phi e^{\tau T_+}
e^{\phi T_-} G_{0,-} G_{0,+} \phi_i(1) |\phi_j \rangle
\end{equation}
where we have inserted $\phi_i$ at the point 1. The notation
$\pm$ refers to light-cone type components and $G$ in (208) is
the superpartner of the energy-momentum tensor. Formally we can
interpret (208) (compare with equation (164)) as a computation in
the topological matter theory coupled to topological gravity.

Let us now assume that in the product $\phi_i \phi_j$ there is
some piece of the type $W'F$.
Using the SUSY transformations of the Landau-Ginzburg
lagrangian, we can write
\begin{equation}
W' F = Q ( \rho_- F)
\end{equation}
with $\rho$ the fermionic zero-form and $Q$ the BRST charge.
Now we realize from the
Landau-Ginzburg representation of $G_{0,-}$ \cite{L,EK} that
\begin{equation}
G_{0,-} (\rho_- F) \neq 0
\end{equation}
moreover, using the conmutation relations between $Q$
and $G_{0,-}$ we get from (208)
\begin{equation}
| C(\phi_i \phi_j) \rangle = \int_{0}^{\infty} d\tau
e^{-\tau T_+} T_+ G_{0,-} | \psi \rangle
\end{equation}
where $Q(\rho_- F) \! \equiv \! Q \psi$. After a finite energy
regularization
we finally obtain
\begin{equation}
| C(\phi_i \phi_j) \rangle = G_{0,-} | \psi \rangle
\end{equation}
as the result from the "sewing"-representation of (208).

The reader should notice that the key step in the derivation is
equation (210), i.e. we have in the product $\phi_i \phi_j$ a
BRST exact state $Q|\psi \rangle $ such that $G_{0,-}|\psi
\rangle \! \neq \! 0$. Using this fact we can define a notion
of equivariant cohomology by the conditions
\begin{eqnarray}
Q | \phi \rangle & = & 0 \\
G_{0,-} | \phi \rangle & = & 0 \nonumber
\end{eqnarray}
and to describe the previous computation by simply saying that
the contribution to $C(\phi_i \phi_j)$ is determined by
non trivial elements in the equivariant cohomology (213).

At this point we can make contact with the equivariant
cohomology of the topological matter theory coupled to
topological gravity. Following \cite{EK}, we will present
a simple example. Let us consider, for all couplings
$t_i\!=\!0$, the first Landau-Ginzburg gravitational descendant
\begin{equation}
\sigma_1 (P) = W'x = Q( \rho_- x)
\end{equation}
This is non trivial in the equivariant cohomology (213). We want
now to compare $\sigma_1(P)$ in representation (214) with the
dilaton of the theory coupled to topological gravity (151)
\begin{equation}
\gamma_0 = \frac{1}{2} {\hat Q} ( \partial c + c \partial \phi -
{\bar \partial} c - c {\bar \partial} \phi )
\end{equation}
with $\hat Q$ the $N\!=\!2$ BRST generator of the coupled
theory. The equivariant cohomology condition for the coupled
theory is defined by the condition
\begin{equation}
(b_0 + G_0 )_- | \psi \rangle = 0
\end{equation}
with $G_0$ the total super energy momentum tensor. It is now
easy to check that
\begin{equation}
(b_0 + G_0 )_- (\rho_- x + \partial c + c \partial \phi - {\bar
\partial} c - c {\bar \partial} \phi ) = 0
\end{equation}
In summary, there exists a map between the "matter" equivariant
cohomology defined by (213) and the equivariant cohomology of the
topological string obtained after coupling topological matter to
topological gravity.

\vspace{3mm}

{\bf Comment on the physical meaning of equivariant cohomology
in string theory}

The physical motivation for the requirement of equivariant
cohomology, comes from the operator formalism definition of
string amplitudes. As it is well known \cite{OF}, string amplitudes
are
defined by associating with a Riemann surface $\Sigma_{g,s}$
equiped with a set $\{ \xi_s \}$ of local coordinates around the
punctures, a state $|\Sigma_{g,s} \{ \xi_s \} \rangle \in
\otimes^s {\cal H}$ with $\cal H$ the Hilbert space of the
matter and ghost system. On $\cal H$ it is defined a nilpotent
BRST operator $Q$. Physical amplitudes for a set of s-external
states $|\chi_i \rangle$ are defined by
\begin{equation}
\int_{M_{g,s}} \langle \chi_1 | ... \langle \chi_s | \prod_{i,
{\bar i} =1}^{3g-3} b(\chi_i) {\bar b}({\bar \chi}_{\bar i})
\prod_{j,{\bar j}=1}^{s} \oint b(V_j) {\bar b}( {\bar V}_{\bar
j}) |\Sigma_{g,s} \{ \xi_s \} \rangle
\end{equation}
where $V_j$, ${\bar V}_{\bar j}$ are vector fields over the
Riemann surface. To (218) we should imposse two requirements

i) \hspace{5mm}   reparametrization invariance

ii) \hspace{3mm} BRST-invariance

The condition i) implies that (218) should be invariant under any
change of local coordinates. This means in particular invariance
under change $V_i \rightarrow V_i + \delta V_i$ with $\delta
V_i$ a vector field that extends holomorphically inside the disc
around the puncture and it is zero at the puncture, i.e. a
"vertical" vector field refered to the bundle ${\hat M}_{g,s}
\rightarrow M_{g,s}$ with ${\hat M}_{g,s}$ the augmented moduli
space.
The requirements i) and ii) are satisfied if we imposse the
equivariant cohomology condition on external states \cite{ND}
\begin{eqnarray}
Q |\chi \rangle & = & 0 \\
b_{0,-} | \chi \rangle & = & 0 \nonumber
\end{eqnarray}

In abstract terms, the ingredients to define the string amplitudes
are a couple $(Q,b)$ such that
\begin{equation}
Q^2  =  0
\end{equation}
\[ \{ Q, b \}  =  T \]
for $T$ the total energy-momentum tensor and the physical
spectrum being defined by the equivariant cohomology (219).

In standard string theory $(Q,b)$ are respectively identified
with the BRST charge and the $b$-ghost. However it is in
principle possible to define formally generalizations of string
theory where $(Q,b)$ are more general solutions to (219). One
particular case that we will discuss later is to use the
$N\!=\!2$ SUSY pair (1) $(Q^+, Q^-)$.

\vspace{1cm}

\subsection{Gravity and the $t$-part of the $t{\bar t}$-equations}

\vspace{7mm}

A natural way to think about the geometry of the space of TFT's
is as an indirect description of the topological matter coupled
to topological gravity (see for instance \cite{L}).
The logic for this point of view is
that any connection in the space of theories should be
determined by integrating fields, which already implies to
construct forms on the moduli space of the Riemann surface. This
is certainly a fruitful approach at least when we work at genus
zero. In this spirit we can easily derive the $t$-part of the
$t{\bar t}$-equations from two very natural string postulates
\begin{eqnarray}
{\textstyle p.1)} \hspace{18mm} \langle \phi_{i_1} \phi_{i_2}...
\int \phi_{i_s}^{(2)} \rangle & = & \langle \int \phi_{i_1}^{(2)}
\phi_{i_2} ... \phi_{i_s} \rangle \\
{\textstyle p.2)} \hspace{5mm} \langle \phi_{i_1} \phi_{i_2}...
\int \phi_{i_{s-1}}^{(2)} \int \phi_{i_s}^{(2)} \rangle & = &
\langle \phi_{i_1} \phi_{i_2}... \int \phi_{i_s}^{(2)} \int
\phi_{i_{s-1}}^{(2)} \rangle \nonumber
\end{eqnarray}
Defining
\begin{equation}
\langle \phi_i \phi_j \phi_k \int \phi_{l}^{(2)} \rangle \equiv
\partial_l \langle \phi_i \phi_j \phi_k {\textstyle exp} ( {\cal
L}_0 \! + \! \sum_a \! t_a \! \int \! \! \phi_{a}^{(2)} ) \rangle
\end{equation}
we will get from (221)
\begin{equation}
[ \partial_i , C_j ]  =  0
\end{equation}
\[ [ \partial_i ,\partial_j ]  =  0 \]
which are the $t$-part of the $t{\bar t}$-equations (see
equation 58). Now
we can try to use p.1) and p.2) as constrains on the
Landau-Ginzburg description of the TFT ${\cal L} \!=\! {\cal
L}_0 \!+\! \sum_i t_i \int \phi_{i}^{(2)}$. In fact assuming the
existence of a superpotential $W(X,t)$ and some polynomial
representation $\phi_i (X,t)$ of the chiral fields $\phi_i$,
and using the residue formulae representation
\begin{equation}
\langle \phi_i \phi_j \phi_k \int
\phi_{l}^{(2)} \rangle =
\frac{\partial}{\partial t_l} \int \frac{\phi_i \phi_j
\phi_k}{W'} dX
\end{equation}
we can ask ourselves how much information we get from the string
postulates (221), and moreover if the Landau-Ginzburg
representation (224) satisfies them in a natural way.
In fact, this is the case for
\begin{eqnarray}
\phi_i & = & \frac{\partial W}{\partial t_i} \\
\frac{\partial \phi_i}{\partial t_j} & = &
C( \phi_i, \phi_j) \nonumber
\end{eqnarray}
with $C(\phi_i, \phi_j) \!=\! C(\phi_j, \phi_i)$, the symmetric
contact terms defined in the previous section.
The natural questions now will be

a) \hspace{5mm} To get a string, i.e. gravitational,
interpretation of the $t{\bar t}$-part of the
$t{\bar t}$-equations.

b) \hspace{5mm} To use the $t{\bar t}$-equations as a way to
find, at least partially, the Landau-Ginzburg description of
more general lagrangians with ${\bar t}_i$-couplings different
from zero.

\vspace{1cm}

\subsection{Verlinde-Verlinde Contact Term Algebra [20]:
Pure Topological Gravity}

\vspace{7mm}
Saito's recursion relation (196) can be derived, using again
cancel propagator arguments, in the context of pure topological
gravity. From (208) we get
\begin{equation}
| C(P, \sigma_n) \rangle = \int_{0}^{\infty} d \tau
\int_{0}^{2\pi}
d \phi {\textstyle e}^{\tau
T_+} {\textstyle e}^{\phi T_-} G_{0,-} G_{0,+} P(1) |
\sigma_n \rangle
\end{equation}
Using the insertions of $b_0$, ${\bar b}_0$ and
$\delta(\beta_0)$, $\delta ({\bar \beta}_0)$ associated to the
moduli of each puncture, we can set $P\!=\!1$
\begin{equation}
\int_{0}^{\infty} d \tau \int_{0}^{2 \pi} d \phi
{\textstyle e}^{ \tau T_+}
{\textstyle e}^{\phi T_-} G_{0,-} G_{0,+}
| \sigma_n \rangle
\end{equation}
{}From the representation
\begin{equation}
| \sigma_n \rangle = \gamma_0 | \sigma_{n-1} \rangle
\end{equation}
equation (215), and the operator product expansion
\begin{equation}
T(z) \phi (w) = \frac{1}{(z-w)^2} + \frac{\partial \phi}{(z-w)}
\end{equation}
we reobtain Saito's recursion relatio (196)
\begin{equation}
| C(P, \sigma_n) \rangle = | \sigma_{n-1} \rangle
\end{equation}
Using now
\begin{equation}
\sigma_n = \sigma_{n}^{(0)} P
\end{equation}
we can define\footnote{We fix $\sigma_n$ at the point zero and
only integrate the puncture operator.}
\begin{equation}
|C(\sigma_n, \sigma_m) \rangle = \sigma_{n}^{(0)} \int_D
P^{(2)} | \sigma_m \rangle =  | \sigma_{n+m-1} \rangle
\end{equation}
for the rest of the contact terms. Notice that the structure of
the contact terms (232) is consistent with ghost number
conservation
\begin{equation}
gh ( \int \sigma_{n}^{(2)} | \sigma_m \rangle ) = (n-1)+m =
gh ( | \sigma_{n+m-1} \rangle )
\end{equation}

In the derivation of (230) and (232) we have not included the
curvature factor (154). In order to include these contributions,
we can use the following trick. Let us consider the correlator
$\langle \sigma_{n_1} ... \sigma_{n_s} \rangle_g$ for all
$t_i\!=\!0$\footnote{Here the $t_i$ are the couplings in pure
topological gravity} and satisfying
\begin{equation}
\sum_{i=1}^{s} (n_i -1) = 3g-3
\end{equation}
In terms of the string coupling constant $\lambda$ we know that
$\langle \sigma_{n_1} ... \sigma_{n_s} \rangle_g$ goes like
$\lambda^{2g\!-\!2\!+\!n}$ with $(2g\!-\!2\!+\!n)$ the number of
3-vertex necessary for the sewing construction of
$\Sigma_{g,s}$. From recursion relations and the puncture
equation (177) we have \cite{DW}, for all $t_i\!=\!0$, that
\begin{equation}
\lambda \frac{\partial}{\partial \lambda} =
\frac{\partial}{\partial t_1} \rightarrow \sigma_1
\end{equation}
and the dilaton equation
\begin{equation}
\langle \sigma_1 \sigma_{n_1}...\sigma_{n_s} \rangle_g =
(2g-2+n) \langle \sigma_{n_1} ... \sigma_{n_s} \rangle_g
\end{equation}
Defining
\begin{equation}
{\hat \sigma}_n = {\textstyle e}^{\frac{2}{3}(n-1)\pi}\sigma_n
\end{equation}
where the $\pi$ is the conjugate of the Liouville field, we
localize the curvature at the insertion points. Therefore we
expect to derive (236) exclusively from the contribution of
contact terms
\begin{equation}
\langle {\hat \sigma}_1 {\hat \sigma}_{n_1}...{\hat
\sigma}_{n_s} \rangle_g = \sum_{i=1}^{s} C({\hat \sigma}_1,
{\hat \sigma}_{n_i}) \langle {\hat \sigma}_{n_1}... {\hat
\sigma}_{n_s} \rangle_g = (2g-2+n) \langle {\hat \sigma}_{n_1}...
{\hat \sigma}_{n_s} \rangle_g
\end{equation}
{}From (234) we obtain
\begin{equation}
| C({\hat \sigma}_1,{\hat \sigma}_n ) \rangle = \frac{1}{3}
(2n+1) |{\hat \sigma}_n \rangle
\end{equation}

In general the contact term algebra will be given by
\begin{equation}
\int_D {\hat \sigma}_{n}^{(2)} | {\hat \sigma}_m \rangle =
A_{n}^{m} |
{\hat \sigma}_{n+m-1} \rangle
\end{equation}
for certain coefficients $A_{n}^{m}$. From now on we will omit
the superindex (2) in the expression of contact terms (see (240)),
in order to simplify notation.
Assuming representation
(232), we obtain that $A_{n}^{m}$ will depend only on $m$
\begin{equation}
A_{n}^{m} \equiv A_m
\end{equation}
with $A_m$ defined by
\begin{equation}
\int_D {\hat P} | {\hat \sigma}_m \rangle = A_m | {\hat
\sigma}_{m-1} \rangle
\end{equation}
{}From (239) we can conclude that
\begin{equation}
A_m = \frac{1}{3} (2m+1)
\end{equation}

To check the previous argument, we should imposse the
consistency conditions of type (221.b)
\begin{equation}
\int_D {\hat \sigma}_{n_1} \int_D {\hat \sigma}_{n_2} | {\hat
\sigma}_{n_3} \rangle = \int_D {\hat \sigma}_{n_2} \int_D {\hat
\sigma}_{n_1} | {\hat \sigma}_{n_3} \rangle
\end{equation}
which implies
\begin{equation}
A_{n_1}^{n_3+n_2-1} A_{n_2}^{n_3} - A_{n_2}^{n_3+n_1-1}
A_{n_1}^{n_3} = C_{n_2 n_1} A_{n_2+n_1-1}^{n_3}
\end{equation}
with
\begin{equation}
C_{n_2 n_1} = A_{n_1}^{n_2} - A_{n_2}^{n_1}
\end{equation}
The coefficients $A_{n}^{m}$ given by (241) and (243) are
clearly a solution with
\begin{equation}
C_{n_2 n_1} = \frac{2}{3} (n_2 -n_1)
\end{equation}
Notice that Saito's recirsion relation (196) becomes now
\begin{equation}
| C({\hat P}, {\hat \sigma}_n) \rangle = \frac{1}{3}
(2n+1) |{\hat \sigma}_{n-1} \rangle
\end{equation}
An interesting exercise will be to derive relation (248)
directly from the Landau-Ginzburg description.

The consistency of the asymmetric contact term algebra with the
string requirement
\begin{equation}
\langle {\hat \sigma}_n {\hat \sigma}_m \prod_{i=1}^{s} {\hat
\sigma}_{n_i} \rangle_g = \langle {\hat \sigma}_m {\hat \sigma}_n
\prod_{i=1}^{s} {\hat \sigma}_{n_i} \rangle_g
\end{equation}
imposse severe constrains on the
correlators. From (240), (243) and (249), we conclude
\begin{equation}
\frac{2}{3} (m-n) \langle {\hat \sigma}_{n+m-1} \prod_{i=1}^{s}
{\hat \sigma}_{n_i} \rangle_g = \sum_{i=1}^{s} {\cal R}_{D_i} +
{\cal R}_{\Delta}
\end{equation}
with ${\cal R}_{D_i}$ the conmutator of the contact terms of
${\hat \sigma}_n$ and ${\hat \sigma}_m$ with the ${\hat
\sigma}_{n_i}$, and ${\cal R}_{\Delta}$ the conmutator for the
node contribution. Equation (250) clearly shows one of the most
important results of pure topological gravity, namely that
correlators are saturated by the contribution of the boundary of
moduli space. The contributions from the nodes ${\cal
R}_{\Delta}$ are of two types. One corresponds to
the pinching of a handle, which results in a correlator at
genus $g\!-\!1$. The other corresponds to factorizations of the
original surface in two of genus $g\!-\!r$ and $r$ respectively.
Therefore, equation (250) originates recursion relations
relating correlators at genus $g$ and $g'\! <  \! g$.
These recursion relations are crucial to show
the equivalence between matrix models and topological strings
\cite{VV,W3,DW}.

\vspace{1cm}

\subsection{The Gravitational Meaning of the
$t{\bar t}$-equations}

\vspace{7mm}

In section 2.5 we have pointed out the equivalence between
moving in the space of theories and coupling to topological
gravity. Using this approach, we formally associate the $t$-part
of the $t{\bar t}$-equations with the string postulates p.1) and
p.2). We can now try to extend our analysis to the whole "$t{\bar
t}$-plane" of topological matter theories, i.e. considering a
generic family of TFT described by
\begin{equation}
{\cal L}= {\cal L}^0 + \sum_i t_i \int \phi_{i}^{(2)} +
\sum_{\bar i} {\bar t}_{\bar i} \int  {\bar \phi}_{\bar i}^{(2)}
\end{equation}
and moving not only in $t$, but also in the $\bar t$ direction.
We will start performing this analysis at genus zero.
Intuitively we should expect to get from this study some extra
information concerning the contribution to topological amplitudes
from the boundary of the moduli space. The logic for this comes
from the standard version of the BRST anomaly in string theory.
Notice that a variation in $\bar t$'s naively implies coupling
to a pure BRST state.

When we change the couplings $t_i$, we are forced to compute
correlators of the type $\langle \phi_{i_1}...\phi_{i_s} \int
\phi_{i}^{(2)}  \rangle$ and we can always interpret the
integration over the position of $\phi_{i}^{(2)}$ in a
gravitational way, associated with the definition of a
Mumford-Morita form on the moduli space of the Riemann sphere
with punctures. When we try to do the same for a variation in
${\bar t}_i$ without changing the way in which we have twisted the
lagrangian ${\cal L}^0$, we inmediatly find some conceptual
problems. The simplest one is the moduli interpretation of the
integration over the position of ${\bar \phi}_{\bar i}^{(2)}$. The
reason for this is that ${\bar \phi}_{\bar i}^{(2)}$, as defined by
(38), involves $Q^+$ and, on the other hand, the integration over
the moduli or sewing parameters in the way described in section
2.4 involves the SUSY charge $Q^-$, i.e.$G_{0,-}$.

The approach we want to present here will consist in
interpreting the variation in the $\bar t$-direction in a
standard gravitational way, in equal footing with the way we
have interpreted the variation in $t$-direction, but at the
price of modifying the field ${\bar \phi}_{\bar i}$. Namely
we introduce a new field ${\hat \phi}_{\bar i}$
by\footnote{The reader should notice an important difference
between the contact term (252) and the one described in (208)
for Landau-Ginzburg models. In the case (208), the couple $(Q,G)$
we use is not Hodge, i.e. $G$ has trivial cohomology, while in
(252) we use for $(Q,G)$ the $N\!=\!2$ SUSY Hodge system, i.e.
$G$ has non trivial cohomology. See more on this phenomena in
section 2.10.}
\begin{equation}
\int_D {\bar \phi}_{\bar i}^{(2)} |\phi_j \rangle \equiv
\int_{0}^{\infty} d \tau \int_{0}^{2 \pi} d \phi {\textstyle
e}^{\tau T_+} {\textstyle e}^{\phi T_-} G_{0,-}^{-}
G_{0,+}^{-} {\hat \phi}_{\bar i} (1) |\phi_j \rangle
\end{equation}
In other words, we use $G_{0}^{-}$ to define the integration over
the insertion ${\bar \phi}_{\bar i}$ and we change the field
${\bar \phi}_{\bar i}$ to ${\hat \phi}_{\bar i}$ in order to take
into account that, in the l.h.s. of (252), ${\bar \phi}_{\bar i}$
was defined by equation (38) in terms of $Q^+$.
In a more compact notation we can write (252) as
\begin{equation}
\int_D {\bar \phi}_{\bar i}^{(2)} | \phi_i \rangle \equiv
| C({\hat \phi}_{\bar i}, \phi_j) \rangle
\end{equation}

In order to characterize the operator ${\hat \phi}_{\bar i}$, we
will use the following constructive path \cite{U}

i) We define a $t{\bar t}$ contact term algebra
including contact terms between topological and antitopological
fields.

ii) We will imposse on this contact term algebra
consistency conditions of the type (244)

iii) From both contact terms in the $t$ and the
$\bar t$ direction, we will try to compute the curvature of the
$t{\bar t}$-"plane", i.e. to derive the $t{\bar t}$-equations.

We have only developped the previous program in the particular
case of ${\hat c}\!=\!3$ theories reducing the $t$ and $\bar t$
deformations to marginal directions, i.e. to the moduli space of
the ${\hat c}\!=\!3$ $N\!=\!2$. However we believe that this
program can be extended to more general cases.

\vspace{1cm}

\subsection{$t{\bar t}$-Contact Term Algebra for ${\hat
c}\!=\!3$ SCFT's}

\vspace{7mm}

Let us consider the algebra of operators generated by:
$\phi_i$, ${\hat \phi}_{\bar i}$ and the dilaton field
$\sigma_{1}$ with $i=1,..,n$ for $n$ the number of marginal
deformations. We define the following contact term
algebra \cite{U}
\begin{eqnarray}
\int_{D} \phi_i \: | \phi_j \rangle \: = \:
\Gamma_{i j}^{k} | \phi_k \rangle
& , & \int_{D} {\hat \phi}_{\bar i} | {\hat \phi}_{\bar j}
\rangle \: = \:
{\tilde \Gamma}_{{\bar i} {\bar j}}^{\bar k}
| {\hat \phi}_{\bar k} \rangle \nonumber \\
\int_{D} {\hat \phi}_{\bar i} | \phi_j \rangle \: = \:
G_{j {\bar i}} | \sigma_{1} \rangle
& , & \int_{D} \phi_i | {\hat \phi}_{\bar j} \rangle \: =  \:
{\widetilde G}_{i {\bar j}} | {\sigma_{1}} \rangle  \nonumber \\
\int_{D} {\sigma}_{1} \: | \phi_i \rangle \: =  \:
a \:  | \phi_i \rangle
& , & \int_{D} \phi_i | {\sigma}_{1} \rangle \: =  \:
b | \phi_i \rangle \\
\int_{D} \sigma_{1} | {\hat \phi}_{\bar i} \rangle \: =  \:
c | {\hat \phi}_{\bar i} \rangle
& , & \int_{D} {\hat \phi}_{\bar i} | {\sigma}_{1} \rangle \: =
\: d | {\hat \phi}_{\bar i} \rangle \nonumber \\
\int_{D} {\sigma}_{1} | {\sigma}_{1} \rangle \: =  \:
e | {\sigma_{1}} \rangle && \nonumber
\end{eqnarray}
\noindent
In order to take into account the contribution of the curvature
and the twist we introduce the operator
\begin{equation}
e^{\frac{1}{2} {\tilde \varphi} (z)}
\end{equation}
for ${\tilde \varphi}\!=\! \varphi \! + \!2 \pi$, where
$\varphi$ is the operator the bosonizes the ${\cal U}(1)$
current of the $N\!=\!2$ SCFT and $\pi$ the conjugate of the
Liouville field.
The contact term algebra for this operator is
defined as follows
\begin{eqnarray}
\int_{D} \phi_i | e^{\frac{1}{2}
{\tilde \varphi} (z)} \rangle \: =  \:
A_{i}  | e^{\frac{1}{2} {\tilde \varphi} (z)} \rangle
& , & \int_{D} {\hat \phi}_{\bar i}
| e^{\frac{1}{2} {\tilde \varphi} (z)} \rangle \: =  \: 0 \\
\int_{D} {\sigma}_{1} | e^{\frac{1}{2} {\tilde \varphi} (z)}
\rangle \: =  \:
a | e^{\frac{1}{2} {\tilde \varphi} (z)} \rangle &  & \nonumber
\end{eqnarray}
\noindent
The undetermined constants appearing in (254) and (256) will be now
fixed by imposing consistency conditions (244).

Before entering into a detailed description of the consistency
conditions
for the algebra (254) (256), we would like to make some comments.
The most
interesting aspect of (254) is the specific $t{\bar t}$-contact
term
\begin{equation}
| C({\hat \phi}_{\bar i}, \phi_j ) \rangle = G_{j{\bar i}}
|\sigma_1 \rangle
\end{equation}
where the topological-antitopological fusion really takes place.
We can argue on (257) in the following way. From the definition
(252) of the field ${\hat \phi}_{\bar i}$ we can formally write
\begin{equation}
{\hat \phi}_{\bar i} \rightarrow Q_{(+)}^+ "\frac{1}{Q_{(-)}^-}"
{\bar \phi}_{\bar i}
\end{equation}
and taking into account the ghost charges of $Q^+$ and $Q^-$ to
interpret ${\hat \phi}_{\bar i}$, at least at the level of ghost
charges, as having implicitely an $n\!=\!2$ gravitational
descendent index. This interpretation as gravitational
descendent should be considered only as an heuristic way to
motivate (257). Even when (258) is purely formal, we
notice that can give a hint on the appearance of the dilaton in
(257), because $"\frac{1}{Q^-}"$ could be interpreted as a
would-be $c$-ghost field\footnote{As a marginal comment we
notice that for type B-models in the case ${\hat c}\!=\!3$, we
can identify $b_{0}^{-}$ with $\partial$ and the BRST charge
with ${\bar \partial}$. Now we can use the fact that $\partial$,
${\bar \partial}$ define a Hodge structure defined by the $(Q^+,
Q^-)$ $N\!=\!2$ algebra. Using Hodge $\partial{\bar
\partial}$-lemma we define $\frac{1}{Q_{(-)}^-}$ as
$\frac{\partial}{\Delta}$ \cite{BCOV}. This is the
basic lemma needed to define the kinetic part of the
Kodaira-Spencer lagrangian.}.

After this general comment we proceed to solve the consistency
conditions. For this, we will assume

i) That $G_{i,\bar j}$ is invertible.

ii) The value of $a$ equals -1. This condition is based on the
way the dilaton field measures the curvature.

iii) The following derivation rules
\begin{eqnarray}
\phi_i \Gamma_{\alpha \beta}^{\gamma}(t,{\bar t})
& = & \partial_{i} \Gamma_{\alpha \beta}^{\gamma}(t,{\bar t})  \\
{\hat \phi}_{\bar i} \Gamma_{\alpha \beta}^{\gamma}(t,{\bar t})
& = &  (-1)^{F( \Gamma_{\alpha \beta}^{\gamma})} \;
\partial_{\bar i} \Gamma_{\alpha \beta}^{\gamma}(t,{\bar t})
\hspace {5mm}, \hspace{8mm}
F(\Gamma_{\alpha \beta}^{\gamma}) \! =  \! q_{\gamma} \!
- \! q_{\alpha}  \! - \! q_{\beta} \nonumber
\end{eqnarray}
\noindent
where $\Gamma_{\alpha \beta}^{\gamma}$ stands for a generic
contact term tensor, $q_{\alpha}$ for the ${\cal U}(1)$ charge
associated to the corresponding field, and
which defines the way the operators act on the coefficients
appearing in the contact term algebra. Notice that in general these
coefficients will depend on the moduli parameters $(t,{\bar t})$.
The logic for this rule is the equivalence
between the insertion of a marginal field and the derivation with
respect to the corresponding moduli parameter. For this reason
we will not associate any derivative with the dilaton field.
The derivation rule (259.b) is forced by the topological
interpretation of the $\bar t$ insertions we are using.
Once we decide to work with the operators
${\hat \phi}_{\bar i}$ and to define the measure using only $G^{-}$
insertions, we must accommodate to this picture the coupling
of the spin connection to the ${\cal U}(1)$ current. Since the
derivation $\partial_{\bar i}$ corresponds to the insertion
of an antitopological field, we need
to change, in the neighborhood of the insertion, the sign of
the coupling of the ${\cal U}(1)$ current to the background gauge
field defined by the spin connection. This fact gives raise
to the factor $(-1)^{F(\Gamma)}$ in (259.b).

Using i), ii) and iii), let us start by analizing the following
consistency condition
\begin{equation}
\int_{D} \sigma_{1} \int_{D} \phi_i | \phi_j
\rangle = \int_{D} \phi_i \int_{D} \sigma_{1}
| \phi_j \rangle \\
\end{equation}
\noindent
Applying the contact term algebra (254), we get
\begin{equation}
b \: \Gamma_{i j}^{k} | \phi_k \rangle -
\Gamma_{i j}^{k} | \phi_k \rangle =
-2 \: \Gamma_{i j}^{k} | \phi_k \rangle \\
\end{equation}
\noindent
which, for a non vanishing $\Gamma_{i j}^{k}$, implies that
\begin{equation}
b= -1 \\
\end{equation}
\noindent
{}From the condition
\begin{equation}
\int_{D} \sigma_{1} \int_{D} \phi_i | \sigma_{1}
\rangle = \int_{D} \phi_i \int_{D} \sigma_{1}
| \sigma_{1} \rangle \\
\end{equation}
\noindent
together with equation (262) and the derivation rules (259),
we obtain
\begin{equation}
\partial_{i} e | \sigma_{1} \rangle - e | \phi_i
\rangle = | \phi_i \rangle \\
\end{equation}
\noindent
being solved by
\begin{equation}
e=-1 \\
\end{equation}
\noindent
To continue the study, we take the condition
\begin{equation}
\int_{D} {\hat \phi}_{\bar i} \int_{D} {\hat \phi}_{\bar j}
| \phi_k \rangle = \int_{D} {\hat \phi}_{\bar j}
\int_{D} {\hat \phi}_{\bar i} | \phi_k \rangle \\
\end{equation}
\noindent
which leads to
\begin{equation}
( \: {\tilde \Gamma}_{{\bar j} {\bar i}}^{\bar l} \: G_{k {\bar l}}
+  \partial_{\bar i} G_{k {\bar j}} \: ) \: | \sigma_{1} \rangle +
d \: G_{k {\bar j}} \: | {\hat \phi}_{\bar i} \rangle =
(\: {\tilde \Gamma}_{{\bar i} {\bar j}}^{\bar l} \: G_{k {\bar l}}
+  \partial_{\bar j} G_{k {\bar i}} \: ) \: | \sigma_{1} \rangle +
d \: G_{k {\bar i}} \: | {\hat \phi}_{\bar j} \rangle  \\
\end{equation}
\noindent
Using that $G_{i {\bar j}}$ is invertible, and for a general
number of marginal deformations, we get from the above equation
\begin{equation}
d=0 \\
\end{equation}
\noindent
Moreover, the consistency condition
\begin{equation}
\int_{D} \sigma_{1} \int_{D} {\hat \phi}_{\bar i} | \phi_i
\rangle = \int_{D} {\hat \phi}_{\bar i} \int_{D} \sigma_{1}
| \phi_i \rangle \\
\end{equation}
\noindent
and equation (268) imply that
\begin{equation}
c=0 \\
\end{equation}
{}From (262), (265), (268) and the consistency condition
\begin{equation}
\int_{D} \phi_i \int_{D} {\hat \phi}_{\bar j}
| \sigma_{1} \rangle = \int_{D} {\hat \phi}_{\bar j}
\int_{D} \phi_i | \sigma_{1} \rangle \\
\end{equation}
\noindent
we get easily
\begin{equation}
{\widetilde G}_{i {\bar j}} = 0 \\
\end{equation}
\noindent
The next conditions we will analyze involve the curvature
operator $e^{\frac{1}{2} {\tilde \varphi} (z)}$
\begin{eqnarray}
\int_{D} \phi_i \int_{D} {\hat \phi}_{\bar j}
| e^{\frac{1}{2} {\tilde \varphi} (z)} \rangle & = &
\int_{D} {\hat \phi}_{\bar j}
\int_{D} \phi_i | e^{\frac{1}{2} {\tilde \varphi} (z)}
\rangle \\
\int_{D} \phi_i \int_{D} \phi_j
| e^{\frac{1}{2} {\tilde \varphi} (z)} \rangle & = &
\int_{D} \phi_j \int_{D} \phi_i
| e^{\frac{1}{2} {\tilde \varphi} (z)} \rangle \nonumber
\end{eqnarray}
\noindent
from which we get, assuming that $\Gamma_{i j}^{k}$ is symmetric
in the lower
indices\footnote{The symmetry of $\Gamma_{i j}^{k}$ will assure
that
$\int_{D} \phi_i \int_{D} \phi_j
| \phi_k \rangle = \int_{D} \phi_j
\int_{D} \phi_i | \phi_k \rangle$ is satisfied.}
\begin{eqnarray}
 G_{i {\bar j}} & = &
 \partial_{\bar j} A_{i} \\
\partial_{i} A_{j} & =  & \partial_{j} A_{i} \nonumber
\end{eqnarray}
\noindent
Equations (274) imply that the metric
$G_{i {\bar j}}$ is K\"ahler, for a certain potential
$K(t,{\bar t})$
\begin{equation}
G_{i {\bar j}} = \partial_{i} \partial_{\bar j} K \\
\end{equation}
\noindent
With this information, we can return to (267) and deduce that the
tensor
${\tilde \Gamma}_{{\bar i} {\bar j}}^{\bar k}$ is symmetric in
the lower indices
\begin{equation}
{\tilde \Gamma}_{{\bar i} {\bar j}}^{\bar k} =
{\tilde \Gamma}_{{\bar j} {\bar i}}^{\bar k} \\
\end{equation}
\noindent
Using now
\begin{equation}
\int_{D} \phi_i \int_{D} {\hat \phi}_{\bar j}
| {\hat \phi}_{\bar k} \rangle  =
\int_{D} {\hat \phi}_{\bar j} \int_{D} \phi_i
| {\hat \phi}_{\bar k} \rangle \\
\end{equation}
\noindent
we obtain that ${\tilde \Gamma}_{{\bar j} {\bar k}}^{\bar l}$
is only function of
the antitopological variables
\begin{equation}
\partial_{i} {\tilde \Gamma}_{{\bar j} {\bar k}}^{\bar l} = 0 \\
\end{equation}
\noindent
Condition (278),
together with $\int_{D} {\hat \phi}_{\bar i} \int_{D}
{\hat \phi}_{\bar j}
| {\hat \phi}_{\bar k} \rangle  =
\int_{D} {\hat \phi}_{\bar j} \int_{D} {\hat \phi}_{\bar i}
| {\hat \phi}_{\bar k} \rangle$
allow to impose a vanishing contact term for antitopological
operators.

To conclude the study of the consistency conditions we will
consider now the relation
\begin{equation}
\int_{D} \phi_i \int_{D} {\hat \phi}_{\bar j}
| \phi_k \rangle = \int_{D} {\hat \phi}_{\bar j}
\int_{D} \phi_i | \phi_k \rangle \\
\end{equation}
\noindent
Using equations (262) and (272), we obtain
\begin{eqnarray}
\int_{D} \phi_i \int_{D} {\hat \phi}_{\bar j}
| \phi_k \rangle & = &
\partial_{i} G_{k {\bar j}} \: | \sigma_{1} \rangle -
G_{k {\bar j}} \: | \phi_i \rangle - G_{i {\bar j}}
\: | \phi_k \rangle  + fact \; terms \\
\int_{D} {\hat \phi}_{\bar j}
\int_{D} \phi_i | \phi_k \rangle & = &
- \partial_{\bar j} \Gamma_{i k}^{l} \: | \phi_l \rangle
+ \Gamma_{i k}^{l} G_{l {\bar j}} | \sigma_{1} \rangle \nonumber
\end{eqnarray}
\noindent

In order to motive the inclusion of factorization terms in (280),
let us notice two facts. The necessity of including
factorization terms at the level of consistency conditions
(244) is already present in the simplest case of contact term
algebra, i.e. in pure gravity. Due to the
asymmetry of the factors $A_{n}^{m}$ (241), it is not possible to
satisfy the relations $\int
{\hat P} \int {\hat \sigma}_n | {\hat P} \rangle \!= \! \int
{\hat \sigma}_n \int {\hat P} | {\hat P} \rangle$ without taking
factorization terms into account. Second, the heuristic argument
(258)
seems to indicate a hidden gravitational descendent index in
the operators ${\hat \phi}_{\bar i}$. Therefore,
and due to the non vanishing correlation
function $C_{i j k}$ at genus zero for three marginal fields,
we should consider the possible existence of factorization terms
associated to the ${\hat \phi}_{\bar j}$ insertions.
We can write them generically as follows
\begin{equation}
fact \; terms = B_{\bar j}^{l n} \, C_{i k n} \: | \phi_l
\rangle \\
\end{equation}
\noindent
{}From equations (279)-(281),
we obtain that the
coefficient $\Gamma_{i j}^{k}$ is the connection for the metric
$G_{i {\bar j}}$, which we already know that is K\"ahler
\begin{equation}
\Gamma_{i j}^{k} = ( \partial_{i} G_{j {\bar l}} )
G^{{\bar l} k} \\
\end{equation}
\noindent
and a $(t,{\bar t})$ type equation
\begin{equation}
\partial_{\bar n} \Gamma_{i j}^{k} = G_{i {\bar n}} \delta_{j}^{k}
+ G_{j {\bar n}} \delta_{i}^{k} -
B_{\bar n}^{m k} C_{i j m} \\
\end{equation}

The tensor $B_{\bar j}^{l n}$ can be derived from the contact
term algebra by the following argument.
Let's consider the consistency condition on a general
string amplitude
\begin{equation}
\langle {\hat \phi}_{\bar i} {\hat \phi}_{\bar j}
\prod_{l=1}^{s} \phi_l \rangle_{g} =
\langle {\hat \phi}_{\bar j} {\hat \phi}_{\bar i}
\prod_{l=1}^{s} \phi_l \rangle_{g} \\
\end{equation}
\noindent
from (254) we get
\begin{equation}
(\, {\tilde \Gamma}_{{\bar i} {\bar j}}^{\bar k} -
{\tilde \Gamma}_{{\bar j} {\bar i}}^{\bar k} \, ) \langle \,
{\hat \phi}_{\bar k} \,
\prod_{l=1}^{s} \phi_l \, \rangle_{g} = \sum_{l=1}^{s}
{\cal R}_{D_{l}} + \sum_{nodes} {\cal R}_{\Delta} \\
\end{equation}
\noindent
where ${\cal R}_{D_{l}}$ denotes the
commutator of the contact terms of
${\hat \phi}_{\bar i}$ and ${\hat \phi}_{\bar j}$ with
$\phi_l$, and ${\cal R}_{\Delta}$ the commutator of those
at the nodes.
Using now the symmetry of
${\tilde \Gamma}_{{\bar i} {\bar j}}^{\bar k}$
in the lower indices (276), we can conclude
\begin{equation}
\sum_{l=1}^{s} {\cal R}_{D_{l}} = \sum_{nodes} {\cal R}_{\Delta} =0
\end{equation}

The contribution at a node associated with the factorization of the
surface, will be
defined by the tensor $B_{\bar j}^{\alpha \beta}$ as follows
\begin{eqnarray}
\langle \, {\hat \phi}_{\bar i} {\hat \phi}_{\bar j}
\, \prod_{l \in S} \phi_l \, \rangle_{g,{\Delta}}  & = &
\sum_{r=0}^{g} \:
\sum_{X \cup Y = S} [ \:
B_{\bar j}^{\alpha \beta} G_{\alpha {\bar i}}
\: \langle \sigma_{1}
\: \prod_{l \in X} \phi_l \rangle_{r} \:
\langle \phi_{\beta} \: \prod_{n \in Y}
\phi_n \rangle_{g-r} +  \nonumber \\
& + & \partial_{\bar i} B_{\bar j}^{\alpha \beta} \:
\langle \, \phi_{\alpha}
\prod_{l \in X} \phi_l \, \rangle_{r}  \:
\langle \phi_{\beta} \: \prod_{n \in Y}
\phi_n \rangle_{g-r} \: ]
\end{eqnarray}
\noindent
where $S$ refers to the set of all punctures, $X$ and $Y$
is a partition of it, and the tensor $B$ can be chosen symmetric
in the upper indices.
Using now (286) we get
\begin{eqnarray}
B_{\bar i}^{\alpha \beta} G_{\alpha {\bar j}} & = &
B_{\bar j}^{\alpha \beta} G_{\alpha {\bar i}}  \\
\partial_{\bar i} B_{\bar j}^{\alpha \beta}
& = & \partial_{\bar j} B_{\bar i}^{\alpha \beta} \nonumber
\end{eqnarray}
\noindent
By an analogous argument, we find from
condition (279) and for a general string amplitude
\begin{equation}
\partial_{i} B_{\bar j}^{\alpha \beta} + B_{\bar j}^{\alpha \gamma}
\Gamma_{i \gamma}^{\beta} + B_{\bar j}^{\gamma \beta}
\Gamma_{i \gamma}^{\alpha} - 2 \partial_{i} K
B_{\bar j}^{\alpha \beta} = 0
\end{equation}
\noindent
Let's define
$B_{\bar j}^{\alpha \beta} = B_{{\bar j} {\bar \alpha} {\bar \beta}}
e^{2K} G^{{\bar \alpha} {\alpha}} G^{{\bar \beta} {\beta}}$. Then,
equations (288) and (289) imply that
$B_{{\bar i}{\bar j}{\bar \beta}}$
is proportional to the three point correlation function
for the antitopological fields.
Substituting this information into equation (283), we obtain
the $(t,{\bar t})$-equation
\begin{equation}
\partial_{\bar n} \Gamma_{i j}^{k} = G_{i {\bar n}} \delta_{j}^{k}
+ G_{j {\bar n}} \delta_{i}^{k} -
{\bar C}_{\bar n}^{m k} C_{i j m} \\
\end{equation}
\noindent
Notice that in order to get the special geometry relation (290)
from the contact term algebra, it was necessary to make use of
the derivation rule (259.b).
{}From (290) we can conclude that the metric
$G_{i {\bar j}}$ is the Zamolodchikov metric for the
marginal deformations, therefore obtaining the special geometry
of the moduli space of $N\!=\!2$, ${\hat c}\!=\!3$ SCFT's
presented in section 1.8.

{}From the previous result we observe that the combined action on
$t$ defined by the contact terms $\Gamma_{ij}^{k}$, and on
$\bar t$ characterized by $G_{i{\bar j}}$,
produces the whole $t{\bar t}$-connection, concluding for the
case ${\hat c}\!=\!3$ the steps i), ii) and iii) of section 2.7.

\vspace{1cm}

\subsection{Holomorphic Anomaly: the Genus Zero Case}

\vspace{7mm}

Let us write the $t{\bar t}$-equation in the condensed way
\begin{equation}
[ D_i , D_{\bar j} ] = - [ C_i , {\bar C}_{\bar j} ]
\end{equation}
If now we interpret $D_i$, $D_{\bar j}$ as defining the motion
in the space of theories
\begin{eqnarray}
C_{i_1...i_s;j}^0 & \equiv & \langle \phi_{i_1} ... \phi_{i_s}
\int \phi_{j}^{(2)} \rangle \equiv D_j \langle \phi_{i_1}...
\phi_{i_s} \rangle \\
C_{i_1...i_s;{\bar j}}^0 & \equiv & \langle \phi_{i_1} ...
\phi_{i_s}
\int {\bar \phi}_{\bar j}^{(2)} \rangle \equiv D_{\bar j}
\langle \phi_{i_1}... \phi_{i_s} \rangle \nonumber
\end{eqnarray}
we get
\begin{equation}
D_{\bar j} C_{i_1 ...i_s ;{\bar i}}^0 = [D_i, D_{\bar j}]
C_{i_1...i_s}^0 + D_i C_{i_1 ... i_s; {\bar j}}^{0}
\end{equation}
and even if we start with holomorphic correlators $C_{i_1 ...
i_s}^0$ for a topological field theory, we will find for the
correlators of a neigborhood theory defined by $C_{i_1...i_s;
i}^0$ an anomalous contribution coming from (291). This anomalous
contribution, first discovered by \cite{BCOV}, is known as
holomorphic anomaly. The physical origin of this anomaly is
associated with the fact that derivatives with respect to the
couplings of pure BRST operators are not any more pure BRST. In
principle the anomaly (293), at least at genus zero, is a general
fact independently of the value of ${\hat c}$. However it is
only for the special case ${\hat c}\!=\!3$ that we can interpret
this anomaly using the tools we have introduced in the previous
section. For ${\hat c}\!=\!3$ and reducing to marginal $t$ and
$\bar t$ deformations, the only non-vanishing correlators at
genus zero are of the form
\begin{equation}
C_{i_1 i_2 i_3; j_1 ... j_s}^0 = \langle \phi_{i_1} \phi_{i_2}
\phi_{i_3} \int \phi_{j_1}^{(2)} ...\int \phi_{j_s}^{(2)}
\rangle_0
\end{equation}
and therefore all of them should define measures on the moduli
space of Riemann surfaces with $n+3$ punctures. In other words,
the study of these correlators is strictly equivalent to couple
the matter theory to topological gravity. Using (292), correlators
can be expressed in terms of the three point
functions
\begin{equation}
C_{i_1 i_2 i_3 ; j_1 ... j_s}^0 = D_{j_s} ... D_{j_1} C_{i_1 i_2
i_3}^0
\end{equation}
Their anomalous piece can be computed applying
succesive times equation (293) togheter with the $t{\bar
t}$-equation (291). In particular the $t{\bar t}$-equation can be
seen as the simplest case of the holomorphic anomaly, i.e. for
the four point function $C_{i_1 i_2 i_3 ; j}$.

\vspace{1cm}

\subsection{Higher Genus and Quantum Geometry}

\vspace{7mm}

Until now we have reduced our discussion to the case of
genus zero. It is in this reduced framework where we have
connected the geometry of the space of theories with the physics
of topological strings. Once we have topological matter coupled
to topological gravity, nothing prevent us a priori for
computing higher genus amplitudes. It is on the basis of these
amplitudes that some form of quantum geometry should appear in
the future.

One of the more important facts we have observed in the study of
pure topological gravity are the recursion relations, by means
of which we can construct genus $g$ amplitudes in terms of genus
$(g\!-\!1)$ amplitudes. The origin of these recursion relations
is the zero contribution from the bulk. It would be certainly
important to generalize these type of recursion relations to
generic topological strings. A way to begin this project,
initiated in \cite{BCOV}, is to generalize the holomorphic anomaly
to higher genus amplitudes.

For the case ${\hat c}\!=\!3$ a generic correlator
$C_{i_1...i_s}^g$ for marginal fields at genus $g$ is defined by
\begin{equation}
C_{i_1...i_s}^g = \int_{M_{g,s+1}} \langle
\oint_{C_{z_1}} \! \! G^- {\bar G}^- \phi_{i_1}
... \oint_{C_{z_s}} \! \! G^- {\bar G}^- \phi_{i_s}
\prod_{j,{\bar j}=1}^{3g-3} G^- (\chi_j)
{\bar G}^- ({\bar \chi}_{\bar j}) \rangle
\end{equation}
with $\chi_j$, ${\bar \chi}_{\bar j}$ the Beltrami
differentials and where $\oint_{C_{z_i}} G^- {\bar G}^- \phi_i
\! = \! \phi_{i}^{(2)}$ (see equation (36)).
The correlator (296) have the same structure as a
correlator in the bosonic string provided we interpret the
$G^-$'s as the $b,{\bar b}$-ghosts. The difference however is
that, as we have already mention in section 2.1, the factor
$\prod_{j,{\bar j}=1}^{3g-3} G^- (\chi_j)
{\bar G}^- ({\bar \chi}_{\bar j})$ is
coming from the integration over the supermoduli and therefore
we are forced a priori to define the string measures using a
pair $(Q,b)$ which in addition to the standard
requirement $\{ Q,  b \} \!=\!T$ defines Hodge structure,
i.e. the cohomology of $Q$ is isomorphic to the
cohomology of the field $b$.
We have already feel this fact in the
computations at genus zero in the definition of contact terms.
{}From a physical point of view the first implication of defining
string amplitudes using a $(Q, b)$ system which at the
same time satisfies the $N\!=\!2$ algebra, i.e. it is Hodge,
is that the propagators
\begin{equation}
\frac{ b_{0,+} b_{0,-}}{L_0+{\bar L}_0}
\end{equation}
which we are going to associate with the sewing operators in
order to define the string amplitudes, project out all zero
energy states. These simple reasoning seems a priori to prevent
any consistent way to define genus $g$ amplitudes for external
zero energy states in terms of amplitudes at genus $(g\!-\!1)$
for again external zero energy states. The
holomorphic anomaly can be extended to genus $g$
amplitudes and partially solves this puzzle.

We will derive the anomaly for correlators at any genus $g$
using again the contact term algebra introduced in section 2.8.
Let us remember the expression of the $t{\bar t}$-amplitudes
with the help of the formal operators ${\hat \phi}_{\bar j}$
\begin{eqnarray}
& \partial_{\bar t} & C_{i_{1} \ldots i_{s}}^{g}  =  \nonumber \\
& = & \int_{{\cal M}_{g,s+1}}  \langle \;
\oint_{C_z} \! \! G^{-} \bar{G}^{-} {\hat \phi}_{\bar t}
\prod_{i=1}^{s} \oint_{C_{z_i}} \! \! G^- {\bar G}^-
\phi_{i} \; \prod_{a,{\bar a }=1}^{3g-3}
\: G^{-}(\chi_a)
\bar{G}^{-} (\bar{\chi}_{\bar a}) \; \rangle_{\Sigma_{g,s+1}} =
\nonumber \\
& = & \langle \: {\hat \phi}_{\bar t} \prod_{i \in S}
\phi_i \: \rangle_{g}
\end{eqnarray}
\noindent
where $S$ notes the set of all punctures and
we have introduced the last equality to simplify the notation.
The contributions to (298) can be written:
\begin{equation}
\langle \: {\hat \phi}_{\bar t} \prod_{i \in S}
\phi_i \: \rangle_{g} = \sum_{i \in S} R_{D_{i}}
+ \sum_{nodes} R_{\Delta} \\
\end{equation}
\noindent
where $R_{D_{i}}$ is the contact term of
${\hat \phi}_{\bar t}$ with the $\phi_i$ insertion, and
$R_{\Delta}$ the contact term contribution that factorize the
surface through a node.
Let's start by analyzing the $R_{D_{i}}$ boundaries:
\begin{eqnarray}
\sum_{i \in S} R_{D_{i}} & = & \sum_{i \in S} \: \langle \:
{\hat \phi}_{\bar t} \prod_{j \in S} \phi_j \:
\rangle_{D_{i}} \: = \: \sum_{i \in S} \:
G_{i {\bar t}} \: \langle \: \sigma_{1} \prod_{j \neq i}
\phi_j \: \rangle = \: \nonumber \\
& = & \sum_{i \in S} \: G_{i {\bar t}} \:
(2 \! - \! 2g \! - \! s \! + \! 1) \: \langle
\: \prod_{j \neq i} \phi_j \: \rangle
\end{eqnarray}
\noindent
The internal nodes $\Delta$ are associated to the
two types of boundaries of a Riemann surface of genus
$g$ and $s$ punctures.
The first one, we will note it as $\Delta_{1}$,
comes from pinching a handle, leading to a surface of
genus $g \! - \! 1$:
\begin{equation}
\langle \: {\hat \phi}_{\bar t} \prod_{i \in S} \phi_i \:
\rangle_{g, \: {\Delta_{1}}} \; = \; \frac{1}{2} \:
B_{\bar t}^{' \alpha \beta}
\langle \: \phi_{\alpha} \phi_{\beta}
\prod_{i \in S} \phi_i \: \rangle_{g-1} \\
\end{equation}
\noindent
where the factor $\frac{1}{2}$ should be added
to reflect the equivalency between the order
in which the two new insertions $\phi_{\alpha}$
are integrated.
The factorization tensor $B'$ satisfies the same set of
equations (288) and (289) that the tensor $B$,
thus it is also proportional to the three point correlation
function.
With an appropriate choice of normalization of the string
amplitudes,
the proportionality constant between both factorization
tensors can be set equal to one \cite{Li}.

The second ones, noted $\Delta_{2}$,
come from the factorization of the surface into
two surfaces of genus $r$ and punctures in the subset $X$,
and genus $g\!-\!r$ and punctures in $Y$ respectively:
\begin{equation}
\langle \: {\hat \phi}_{\bar t} \prod_{i \in S} \phi_i \:
\rangle_{g, \: {\Delta_{2}}} \; = \; \frac{1}{2} \:
\sum_{r=0}^{g} \:
\sum_{X \cup Y = S}
{\bar C}_{\bar t}^{\alpha \beta} \: \langle \phi_{\alpha}
\: \prod_{j \in X} \phi_j \rangle_{r} \:
\langle \phi_{\beta} \: \prod_{k \in Y}
\phi_k \rangle_{g-r} \\
\end{equation}
\noindent
Collecting now equations (300), (301) and (302), we obtain the
equation for the $\bar t$-dependence of any string amplitude:
\begin{eqnarray}
\partial_{\bar t} \: \langle \: \prod_{i \in S} \phi_i
\: \rangle_{g}
& = & \frac{1}{2} \: {\bar C}_{\bar t}^{\alpha \beta} \:
\langle \: \phi_{\alpha} \phi_{\beta}
\prod_{i \in S} \phi_i \: \rangle_{g-1} + \nonumber \\
& + & \frac{1}{2} \: \sum_{r=0}^{g} \:
\sum_{X \cup Y = S}
{\bar C}_{\bar t}^{\alpha \beta} \: \langle \: \phi_{\alpha}
\: \prod_{j \in X} \phi_j \: \rangle_{r} \:
\langle \: \phi_{\beta} \: \prod_{k \in Y}
\phi_k \: \rangle_{g-r} \; + \\
& + & \sum_{i \in S} \: G_{i {\bar t}}\:
(2 \! - \! 2g \! - \! s \! + \! 1) \: \langle
\: \prod_{j \neq i} \phi_j \: \rangle_{g} \nonumber
\end{eqnarray}
Notice that in our derivation of the holomorphic anomaly from
the contact term algebra we have only considered the contact
terms of the antitopological operator ${\hat \phi}_{\bar t}$
with the rest of the operators $\phi_i$ but not the
contact terms among the operators  $\phi_i$ themselves.
This is equivalent to define the correlators
$\langle \prod \phi_i \rangle$
by covariant derivatives of the generating functional. There are
however some aspects of the previous derivation that should be
stressed at this point.

1) The correlators $\langle \prod \phi_i \rangle$
for topological operators can not be
determined by the contact term algebra, by contrast to what
happen in topological gravity. In fact from the contact
term algebra we can only get relations of the type:
\begin{equation}
(\, \Gamma_{i j}^{k} -
\Gamma_{j i}^{k} \, ) \langle \,
\phi_k \,
\prod_{l \in S} \phi_l \, \rangle = \sum_{l \in S}
{\cal R}_{D_{l}} + \sum_{nodes} {\cal R}_{\Delta} \\
\end{equation}
\noindent
which does not imply ($\Gamma_{i j}^{k} -
\Gamma_{j i}^{k} = 0$) anything on the surface
contribution. Moreover they are compatible with making all
contact terms $R_{D_{l}}$ equal to zero by covariantization.

2) If in the computation of
$\langle \, {\hat \phi}_{\bar t} \,
\prod_{i \! \in \! S} \phi_i \, \rangle$
we take into account all
contact terms, i.e contact terms between the $\phi_i$
operators, we
will find, as a consequence of the derivation rules (259) and the
$(t,{\bar t})$ equations (290), that the holomorphic anomaly is
cancelled, reflecting the commutativity of ordinary derivatives
$[ \partial_{\bar i} , \partial_{j} ] = 0$.

3) We should say that from the contact term algebra we can not
prove,
at least directly, that the correlators
$\langle \, {\hat \phi}_{\bar t} \,
\prod_{i\! \in \! S} \phi_i \, \rangle$
are saturated by
contact terms. The fact we have proved is that the contact
term contribution dictated by the contact term algebra
(254) (256) is
precisely the holomorphic anomaly.

4) The curvature of the initial surface is augmented by
two units in both
processes of pinching a handle or factorizing the surface. In order
to take this into account, the two insertions $\phi_{\alpha},
\phi_{\beta}$
generated in these processes should include, in addition, an
extra unit of curvature. Therefore, the total balance of curvature
for the new insertions is zero. This can be seen as the reason for
the zero contact term between the dilaton field $\sigma_{1}$
and the antitopological operators ${\hat \phi}_{\bar i}$ (see
equations (268) and (270)).

To finish this section, we will notice an important property of
the holorphic anomaly equation (303).
Using the covariant prepotential $S$, $S^i\! = \! G^{{\bar i}
i}\partial_{\bar i} S$, $S^{ij}\!=\! G^{{\bar i} i}
\partial_{\bar i} S^j$,
introduced in section 1.8,
we can integrate (303) \cite{BCOV}. From this we get in particular
a Feynman
diagram description of part of the boundary contributions to
$C_{i_1...i_s}^g$.
Let us stress the appearance in the Feynman rules of
a new field, with the defining properties of the dilaton.
This fact has the origin in the pieces depending linearly on the
curvature of the Riemann surface in the expression of the
holomorphic anomaly (303), or from the point of view of the
algebra (254), in the contact term between a topological and a
antitopological field (257).

It would be interesting at this point to reinterpret the Feynman
propagator $S^{ij}$ as a regularization of the Kodaira-Spencer
propagator $\frac{{\bar \partial}^{\dagger} \partial}{\Delta}$
and to connect this regularization with an effective and
operative way for reproducing, using the cancel propagator
argument, the contact term algebra.

\vspace{1cm}

\subsection{Final Comments}

\vspace{7mm}

In this section we will collect some concrete questions which we
believe would be worth to consider in more detail.

i) A direct derivation, using cancel propagator
arguments, of the $t{\bar t}$-connection.

ii) To find a Landau-Ginzburg description of
topological matter theories with $t$ and $\bar t$ couplings
different from zero.

iii) A direct derivation of the renormalization
group "$\beta$-functions" $t_i ( \beta)$, ${\bar t}_{\bar i}
(\beta)$ for $\beta$ the world-sheet scale.

iv) To extend the holomorphic anomaly for correlators
involving gravitational descendents and to massive topological
field theories.

v) To find an effective regularization of the
"Kodaira-Spencer" propagator $\frac{{\bar \partial}^{\dagger}
\partial}{\Delta}$ in a way consistent with the holomorphic
anomaly.

vi) Based on the connection between $t{\bar
t}$-equation and the thermodinamic Bethe ansatz (TBA)
\cite{CVV}, namely TBA as integral representation of
$t{\bar t}$-equations for massive models, to study, from the
$t{\bar t}$-geometry, the integrability of the corresponding
solitonic infrared theory.

vii) To study in a more systematic way properties
of strings defined for a pair $(Q, b)$ which satisfy Hodge
relations, i.e. strings with non-trivial $b$-cohomology.

\vspace{7mm}

{\bf Acknowledgments}

We would like to thank I. Krichever and A. Losev for many
valuable discussions.
This work was partially supported by
grant PB 92-1092, the work of C.G. by Swiss National Science
Foundation and by OFES: contract number 93.0083, and the work
of E.L. by M.E.C. fellowship AP9134090983.

\end{document}